\documentclass[structabstract]{aa}  
\usepackage{graphicx}
\usepackage{txfonts}
 \usepackage{amssymb}
 \usepackage{natbib}
  \usepackage{lscape}
  \bibpunct{(}{)}{;}{a}{,}{,}
  \usepackage{times}

  \usepackage{longtable}
\usepackage{longtable,lscape}
  \usepackage{lscape}
\begin{document}
   \title{The Gaia-ESO Survey: membership and Initial Mass Function of the $\gamma$ Velorum  cluster\thanks{Based on
   observations made with the ESO/VLT, at Paranal Observatory, 
   under program 188.B-3002 (The Gaia-ESO Public Spectroscopic Survey)  }}

   \subtitle{ }
  \author{L. Prisinzano\inst{1}
          \and
          F. Damiani\inst{1}
          \and
          G. Micela\inst{1}
          \and
	  R. D. Jeffries\inst{2}
        \and 
         E. Franciosini\inst{3}
          \and
         G. G. Sacco\inst{3}
          \and       
          A. Frasca\inst{4}
          \and
	  A. Klutsch\inst{4}
	  \and
          A. Lanzafame\inst{4,5}
          \and 
          E. J. Alfaro\inst{6}         
	  \and
	  K. Biazzo\inst{4}
	  \and
          R. Bonito\inst{7}
          \and 
          A. Bragaglia\inst{8}
         \and
           M. Caramazza\inst{1}
           \and
          A. Vallenari\inst{9}
	    \and
	  G. Carraro\inst{10}
        \and
           M. T. Costado\inst{6}
           \and
          E. Flaccomio\inst{1}
          \and
           P. Jofr\'e\inst{11} 
          \and
          C. Lardo\inst{12}
          \and 
          L. Monaco\inst{13}
	  \and 
	  L. Morbidelli\inst{3}
	  \and
	  N. Mowlavi\inst{14}
          \and
          E. Pancino\inst{8}
          \and
          S. Randich\inst{3}
          \and
           S. Zaggia\inst{9}
          }
   \institute{INAF - Osservatorio Astronomico di Palermo, Piazza del Parlamento 1, 90134, Palermo, Italy \\ 
              \email{loredana@astropa.inaf.it}
	\and
        Astrophysics Group, Keele University, Keele, Staffordshire ST5 5BG, United Kingdom 
                \and
        INAF - Osservatorio Astrofisico di Arcetri, Largo E. Fermi 5, 50125, Florence, Italy 
	    Florence, Italy
         \and
        INAF - Osservatorio Astrofisico di Catania, via S. Sofia 78, 95123, Catania, Italy 
        \and
        Dipartimento di Fisica e Astronomia, Universit\`a di Catania, via S. Sofia 78, 95123, Catania, Italy 
        \and
        Instituto de Astrof\'{i}sica de Andaluc\'{i}a-CSIC, Apdo. 3004, 18080 Granada, Spain 
         \and
          Dip. di Fisica e Chimica, Universit\`a di Palermo, P.zza del Parlamento 1, I-90134 Palermo, Italy 
         \and
        INAF - Osservatorio Astronomico di Bologna, via Ranzani 1, 40127, Bologna, Italy 
	    \and 
 	INAF - Padova Observatory, Vicolo dell'Osservatorio 5, 35122 Padova, Italy    
        \and
         European Southern Observatory, Alonso de Cordova 3107 Vitacura, Santiago de Chile, Chile 
         \and
         Institute of Astronomy, University of Cambridge, Madingley Road, Cambridge CB3 0HA, United Kingdom 
  	    \and 
         Astrophysics Research Institute, Liverpool John Moores University, 146 Brownlow Hill, Liverpool L3 5RF, United Kingdom 
	\and
	Departamento de Ciencias Fisicas, Universidad Andres Bello, Republica 220, Santiago, Chile 
	\and 
	Department of Astronomy, University of Geneva, 51 chemin des Maillettes, 1290, Versoix, Switzerland 
         }
     \date{Received  / Accepted}

 
  \abstract
   {Understanding the properties of young  open clusters, such as 
   the Initial Mass Function (IMF), star formation history and
   dynamic evolution, is crucial to obtain reliable theoretical predictions of the mechanisms involved in the 
   star formation process.}  
   {We want to obtain a list, as complete as possible, of confirmed members of the young open cluster
 $\gamma$ Velorum, with the aim of deriving general cluster properties such as the IMF.}
  {We used all available spectroscopic membership indicators within the Gaia-ESO public archive
  together with literature photometry
   and X-ray data and, for each method, we derived the most complete list of candidate cluster members.
   Then,  we considered  photometry, gravity and radial velocities as necessary conditions to select a
   subsample of candidates  whose membership was   confirmed 
   by using the lithium and H$\alpha$ lines and  X-rays as youth indicators.
   }
   {We found 242 confirmed and 4 possible cluster members for which we derived   masses using 
   very recent stellar evolutionary models.
   The cluster IMF in the mass range investigated in this study shows a slope of 
   $\alpha=2.6\pm0.5$   for $0.5<M/M_\odot <1.3$ and $\alpha=1.1\pm0.4$ for $0.16<M/M_\odot <0.5$ and 
    is  consistent  with a standard IMF.}
   {The similarity of the IMF of the young population around $\gamma^2 $Vel to that in other star forming regions and the field suggests it may have formed through very similar processes.
}
    \keywords{stars: pre-main sequence -- (Galaxy:) open clusters and associations: individual: $\gamma$ Velorum, stars: formation --
                        stars: luminosity function, mass function -- techniques: radial velocities -- techniques: spectroscopic}

\authorrunning{L. Prisinzano et al.}
\titlerunning{GES: Membership and IMF of the $\gamma$ Velorum  cluster}
   \maketitle
%

\section{Introduction}
The $\gamma$ Velorum cluster hosts a population of 5-10\,Myr old pre-main sequence (PMS) stars,
located at 356$\pm$11\,pc \citep{jeff09}. Due to its relatively small distance,
it appears quite dispersed on the sky. It does not show  evidence of ongoing star formation
and thus it is an ideal target for studies of young stars in which the accretion phenomena
already have almost entirely ceased \citep{hern08}. The most massive member is $\gamma^2$ Velorum,
a   binary system formed by a Wolf-Rayet (WC8) component of $\sim9\pm2$\,M$_\odot$  and an OIII star
of $30\pm2$\,M$_\odot$ \citep{de-m99} whose initial masses were $\sim 35$ and 31\,M$_\odot$, respectively
\citep{eldr09}.

Discovered in X-rays by \citet{pozz00}, 
the cluster was established thanks to  its relatively high spatial stellar density around 
 $\gamma^2$ Velorum,  
within a region of about one square degree on the sky. 
A deep photometric survey of this cluster has been obtained by \citet{jeff09}, who also used spectroscopic and 
X-ray data to identify the  photometric cluster sequence. 

The  $\gamma$ Velorum cluster was the first  observed in the {\it Gaia}-ESO survey (GES)
\citep{gilm12}, which is a high-resolution spectroscopic survey
using the FLAMES instruments (both GIRAFFE and UVES) of the ESO-VLT \citep{pasq02},
 which aims to obtain a homogeneous overview of the kinematic and chemical abundance distributions 
of several components of our Galaxy, including a  census of  $\sim$100 open clusters  (OCs). 
In particular, the GES observation strategy for the OCs is 
to observe with  GIRAFFE all candidate members falling spatially in the  cluster area and
within the cluster locus of the color-magnitude diagrams (CMD), down to V=19 mag.
The aim of this strategy is to observe an unbiased and
inclusive sample of candidate cluster members.
This observation strategy is adopted to achieve the GES main goals that are to kinematically characterize 
the entire populations, and, at the same time,  
homogeneously derive their chemical abundances. For example, a slightly subsolar metallicity was found by
 \citet{spin14}
for the $\gamma$ Velorum cluster. 
GES data allow also to perform further investigations,   for example to derive 
  fundamental stellar astrophysical parameters  and then cluster fundamental parameters, 
  such as reddening, age, distance and mass. These latter are crucial to
constrain cluster formation theory (star burst events, sequential star formation and age spread),
stellar evolution models and to derive the Initial Mass Function (IMF). 

The first goal of this paper is to establish the membership of the $\gamma$ Velorum  cluster.
Starting from an inclusive sample of candidate cluster members, 
membership will be  confirmed or rejected by using  radial velocities (RV) and  
stellar  properties (e.g., surface gravity, effective temperature, Li abundance,
accretion rates, chromospheric activity, rotation)
that can be derived from  spectral features falling 
in the  $\lambda\lambda 6440-−6815\,\AA$ spectral range, covered by the GIRAFFE HR15N set-up. 
The sample of confirmed members is used to derive the IMF.

In a study dedicated to the dynamical analysis of this cluster, using the very precise RVs 
derived with GES,  \citet{jeff14} found that  the  cluster consists of two distinct kinematic  populations,
referred to as A and B, with ages of about 10\,Myr, of which population B is, on the basis of Li depletion, 
judged to be 1-2\,Myr older than population A.
Since the cluster is located in the region of the Vela OB2 association \citep{de-z99}, 
the authors conclude that   population A is the remnant of an initially much denser cluster,
formed in a denser region of the Vela OB2 association, while  population B is
more extended and supervirial.

This  scenario is coherent with that 
found  by \citet{sacc15} who studied the RV distribution from GES data of the cluster 
NGC\,2547, in the same direction as Vela OB2,
and found an additional population, kinematically distinct from NGC 2547, but
consistent with population B of $\gamma$ Vel  \citep[see also][]{mape15}.

In case of the $\gamma$ Velorum cluster, it is very likely that populations A and B
belong to the same parent nebula,
and, even if the two populations are kinematically distinct, 
they are almost indistinguishable in the CMD and this implies they have very similar distance and ages.
In addition, they share very similar  spectroscopic properties, as already shown in \citet{jeff14}.
For the aims of this work, we thus consider the two populations A and B as a single young population.  



\section{Targets and astrophysical parameters}

%

The GES targets observed in the $\gamma$ Velorum cluster region
  were selected as described in \citet{jeff14}, following the GES
  observational strategy 
 (Bragaglia et al., in preparation).
 
 Candidate cluster members were observed with FLAMES at the VLT using both 
 the GIRAFFE intermediate-resolution and the
 UVES high-resolution spectrographs. Details of the GES observations of the $\gamma$ Velorum cluster 
 are reported in \citet{jeff14}. 
 For our analysis we use only GIRAFFE data while we do not consider UVES data since the sample
 of stars observed with  UVES is not complete, as required for our analysis.
  Data reduction of the GIRAFFE spectra analyzed in this work  has been performed using the 
  pipeline developed at the Cambridge Astronomical Survey Unit (CASU)
  in collaboration with the Keele University,
  as will be described in Lewis et al. (in preparation).
  
  There were 1242 targets observed with GIRAFFE in the  field of $\gamma$ Vel, selected 
  on the basis of their positions in the optical CMDs, but covering a very wide range around the CMD 
  cluster locus.
   Since some targets were observed more than once, the data set includes 1802 spectra. 

The stellar parameters used in this work were taken from the last data release (gesiDR2iDR3) of 
the GES official archive  at the Wide
Field Astronomy Unit (WFAU) of the  Edinburgh University\footnote{http://ges.roe.ac.uk/index.html}.
In particular we used the RVs from the  RecommendedAstroAnalysis table for the 1122 targets 
for which the RVs are given and the RVs from the Spectrum table for the 99 targets
for which the RVs are not given  in the RecommendedAstroAnalysis table. 
The RVs from the Spectrum table were shifted 
by -0.13\,km/s to have the RVs in the same reference system. 
In total we have a RV value for  1221 objects of the entire sample.
The errors on the RV were computed by using the 
RV precision recipe
given in \citet{jack15}. In addition, we used the projected rotational velocities {\it vsin\,i}
from the Spectrum table, while 
the  equivalent width  of the lithium line EW(Li), the full width   at 10\% of the H$\alpha$
peak (H$\alpha$\,10\%), the chromospheric equivalent width  of the H$\alpha$ line and 
the gravity index $\gamma$ (defined in \citet{dami14}), were taken from the  
WgRecommendedAstroAnalysis table \citep{lanz15}.
We also used the $\alpha_c$ index of chromospheric activity based on GES data  from \citet{dami14}.
Finally, we used the optical literature photometry and the EPIC-XMM-Newton X-ray data from \citet{jeff09}.

Double-lined  spectroscopic  binaries  (SB2) were identified  by  examining  the  shape  of  the 
 cross-correlation function while SB1 were classified on the base of  their RV  in case of multiple observations
\citep{lanz15}. 
 In particular, the WgRecommendedAstroAnalysis table of  the $\gamma$ Velorum field includes 
23 SB1 and 21 SB2 stars,  respectively. 

\section{Membership criteria}
We describe here all the adopted criteria used to select  candidate members of the young cluster 
$\gamma$ Velorum. The conditions that we   applied  are all inclusive to select
the maximum number of possible members for each method. This implies the inclusion of a significant fraction
of contaminants, but, as we  describe in Sect.\,\ref{finalmemsection}, 
  the final membership is  based on the necessary conditions from photometry, gravity, 
RV, and an age criterium. The age criterium
is  based on either Li abundance, stellar activity, or X-ray emission, one of those criteria being sufficient.
This strategy ensures the selection of the maximum number of cluster members.

\subsection{Photometric membership\label{photmem}}
\begin{figure}
 \centering
 \includegraphics[width=9cm]{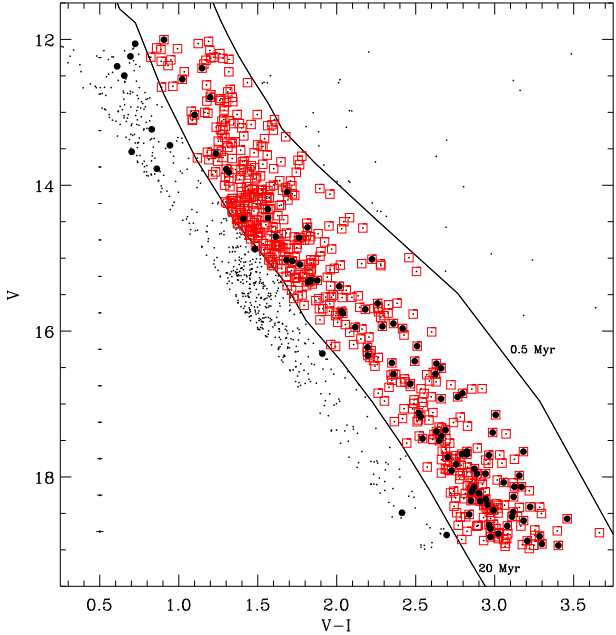}
\caption{Color magnitude diagram of all the 1242 targets observed in the $\gamma$ Velorum  field (dots).
Empty red squares are the 579 
photometric candidate members and black filled circles are X-ray detected objects.
Solid lines are  0.5  and 20\,Myr isochrones from \citet{bara15}. 
Typical photometric error bars are also indicated.}
\label{lithiuma1}
 \end{figure}

As described before,  the survey strategy is to select 
  targets  in a photometric region of the CMD 
larger than that expected for the cluster age. Then in the following analysis we   consider
as high-probability photometric cluster members the 579 objects 
that in the V vs. V-I diagram
fall between the 0.5 and 20\,Myr theoretical isochrones
from 
\citet{bara15}, reddened by E(V-I)=0.055 
and $A_V=0.131$,
at an intrinsic distance modulus of 7.76 \citep{jeff09}, as shown in Fig.\,\ref{lithiuma1}. 
To fix these age limits, we were guided by the position of the X-ray detected objects in the CMD, 
since most of them are expected to be cluster members (see Section\,\ref{xraysection}) 
and thus trace the cluster sequence. With these limits we are confident of including all possible cluster members
but we are aware of including a large fraction of contaminants. However, since we  consider other membership 
criteria, most of the contaminants are discarded in the final selection.

Very young stars with circumstellar disk and/or accretion 
can also be photometrically selected  by considering the IR J-H vs. H-K diagram
where they lie in the well known classical T\,Tauri star (CTTS) locus,
that is a region with IR excesses well outside 
from the locus of the main sequence (MS) or giant stars. This is a way of including additional members, identified by the presence of discs/accretion.
We verified that in this cluster, only 3
of the selected GES targets fall in the CTTS 
locus\footnote{these objects are selected as cluster members with the other methods adopted in this work}  and so we do not consider the IR color-color diagram
as a useful method to select young stars in this cluster.

\subsection{Radial velocities}
\label{radvelsect}
The radial velocity membership criterion is based on the assumption that in a given cluster,
members
 share similar RVs and have a narrow RV distribution. Since our sample of targets has been selected    
photometrically, we expect to find a fraction of contaminant field stars, having a much broader RV
distribution, overlapping with that of the cluster. Our aim is then to model the cluster and field  RV distributions
 to derive the RV range of cluster members.

A scrupulous analysis to model the RV cluster distribution 
has been presented in \citet{jeff14} who considered an unbiased sample of 208  $\gamma$ Velorum members 
and computed, for each member, the likelihood of having the observed RV. This likelihood has been
computed by convolving an intrinsic RV distribution 
with the measurement uncertainties  and the distribution of velocities expected for a given fraction
of binaries. By using a maximum likelihood fit, it has been shown that the cluster RV distribution 
is better represented if the intrinsic RV distribution is modeled with a two-Gaussian fit,
 highlighting the presence of the two kinematic populations A and B in the direction of the $\gamma$ Velorum  cluster.
 
We used the cluster probability density function (PDF) computed  by \citet{jeff14}\footnote{we applied a shift of -0.13\,km/s to
the RVs of the model to move the values to the reference system of the RVs of our data} to derive the
RV range where we can find cluster members. In particular,
by computing the PDF area within a given RV range, 
we fixed the 
RV limits for the cluster to the values for which the probability to find cluster members is smaller 
than 0.003 (equivalent to 3\,$\sigma$ level)
for objects with  RV outside this range. These limits correspond to [$RV_{inf}, RV_{sup}$]=[1.8, 36.5]\,km/s.
The number of cluster members with RVs within this range is 541 while that with 
RVs outside this range is expected to  be 0.3, so this is the best compromise to
 not miss cluster members even though this implies the inclusion of a significant fraction of contaminants.
 We are not considering here the possibility/probability that there is a population of binary systems 
with a broader RV distribution and so  some member binaries may be missed on the basis of their RV.

  In addition, for several aims  of this work,   
  we defined also a more conservative  cluster RV  range corresponding  to a 2\,$\sigma$ confidence level.
  With these conservative RV limits [$RV'_{inf}, RV'_{sup}$]=[12.3, 23.5]\,km/s
we select a less complete  (we expect to miss about 5 cluster members with RV outside these limits)
but less contaminated sample of cluster members that, combined with other conditions, allow us to select
a fiducial sample of almost  certain cluster members. 
  
To compute the contaminant fraction, we fitted the field RV distribution by using 
the entire RV data set but discarding the objects with RVs within the more conservative cluster RV range [$RV'_{inf}, RV'_{sup}$]. 
We   modeled this  field RV distribution with a Gaussian function by using  maximum likelihood fitting and we
found that the RV mean of the field RV distribution  is 54.7$\pm$1.3\,km/s with a $\sigma$=40.2$\pm$0.9\,km/s.

\begin{figure}
 \centering
 \includegraphics[width=9cm]{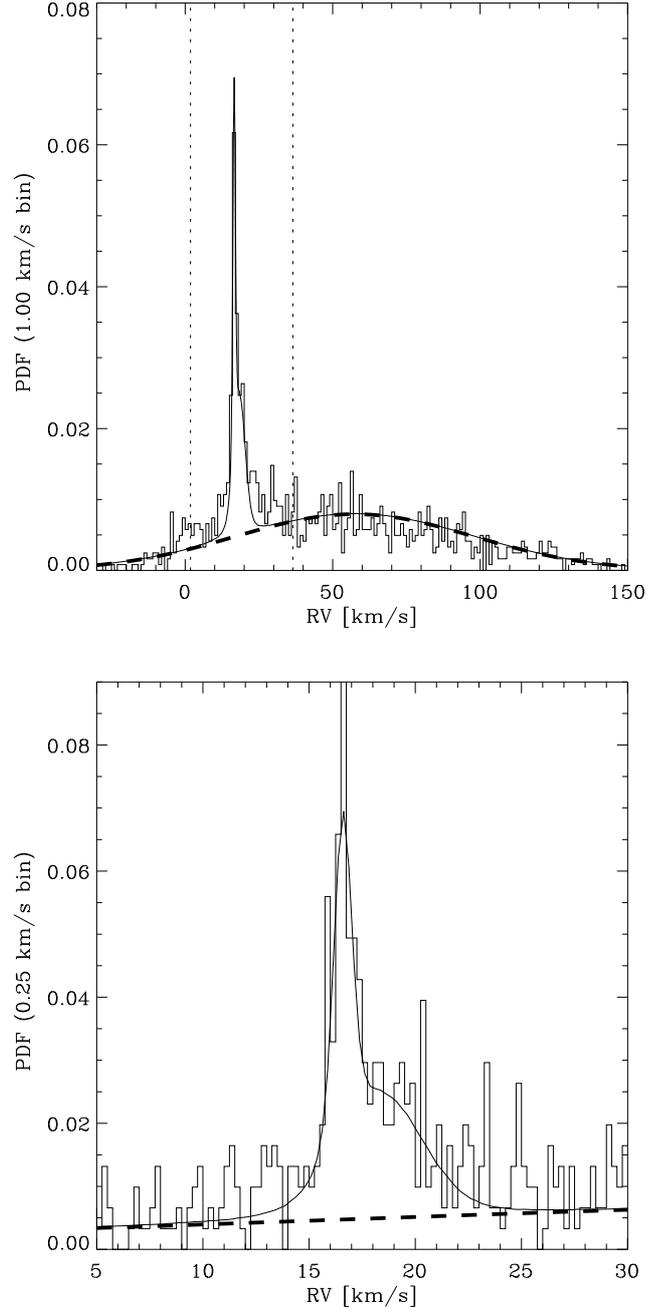}
\caption{The RV histogram for the entire data set of $\gamma$ Velorum cluster 
showing the entire RV range  (upper panel)
and a zoom of the cluster range  (bottom panel) compared with the total PDF (solid line) 
obtained by adding the \citet{jeff14} cluster model to the field PDF performed by us (thick dashed line).
Vertical dotted lines delimits the [$RV_{inf}, RV_{sup}$] range used to select  RV cluster member candidates.}
\label{radvelmem1}
 \end{figure}

Figure\,\ref{radvelmem1} shows the RV density distribution of the entire data set 
 compared to 
 the total PDF obtained by adding the numeric \citet{jeff14} cluster model to  the field PDF derived by us.
The two distributions were normalized to the fraction of objects used to derive the two distributions. 

By using this model, we computed the probability to find field stars within the  cluster [$RV_{inf}, RV_{sup}$] range
and then the number of contaminants expected in the cluster region that amounts to 268 objects. 
We note that the adopted field model  does not accurately describe  our data at $\sim$0\,km/s and  $\sim$30\,km/s,
where there is an excess of stars in the observed distribution. This excess could be due to 
some additional structures in the RV distribution that we do not include in our fit.
For example,  large uncertainties in the RV measurements  of fast rotators can introduce additional structures in the observed 
distribution. This suggests us that the number of contaminants could be larger and so we consider our estimate 
a lower limit to the true contamination. Based on the excess of our data with respect to the model we estimate that the number of
missed contaminats amounts to about 10\%.

In conclusion, we consider candidate members for RV all the 541 stars with RV between   1.8 and 36.5\,km/s.
In this sample we include also the binaries since the RV of the centre of mass  is supposed to share the  cluster RV distribution.
Nevertheless, since the method used to derive official GES released RVs does not ensure  that the RV of the binaries is that
of the centre of mass, we are aware that some 
binary members  may be missed
on the basis of their RV. 
The same is true for fast rotators for which the RV uncertainties are typically very large.
For this reason, binaries and fast rotators are considered as a special sample in the final 
cluster member selection in the sense that for them   RV membership is not considered a necessary condition,
as is, instead,  required for single stars.

\subsection{Lithium line\label{lithiumsection}}

In this section, we assign  cluster membership on the  basis of the 
strength of the Li\,{\tiny I} 6708\,$\AA$ line, that is a well-known age indicator for young stars, 
such as those expected to be found in the $\gamma$ Velorum  cluster.  
As discussed in \citet{jeff14}, theoretical isochrones are very uncertain in predicting 
the lithium depletion pattern and for this reason we adopt an empirical approach 
aimed at highlighting the cluster locus in the EW(Li) vs. V-I diagram
to fix the most appropriate EW(Li) thresholds for the cluster member selection. 
With this aim  we used an initial sample  of candidate cluster members 
based on criteria that are free  from any bias due to the  lithium line.
In particular  we defined a {\it cluster member fiducial sample}
including the 235 objects being both 
photometric cluster members (as defined in Section\,\ref{photmem}) 
and with RV within the conservative cluster range ([$RV'_{inf}, RV'_{sup}$]) defined in the previous section. 
We note that this sample does not include only genuine cluster members since  within the photometric cluster 
locus a fraction of contaminants with RV within the [$RV'_{inf}, RV'_{sup}$] range is expected.
 Nevertheless the sample is strongly dominated by cluster members and can be used to trace
their lithium properties. This sample will be used, as reference for the cluster,
also for other membership criteria described in the following sections.

\begin{figure}
 \centering
 \includegraphics[width=9cm]{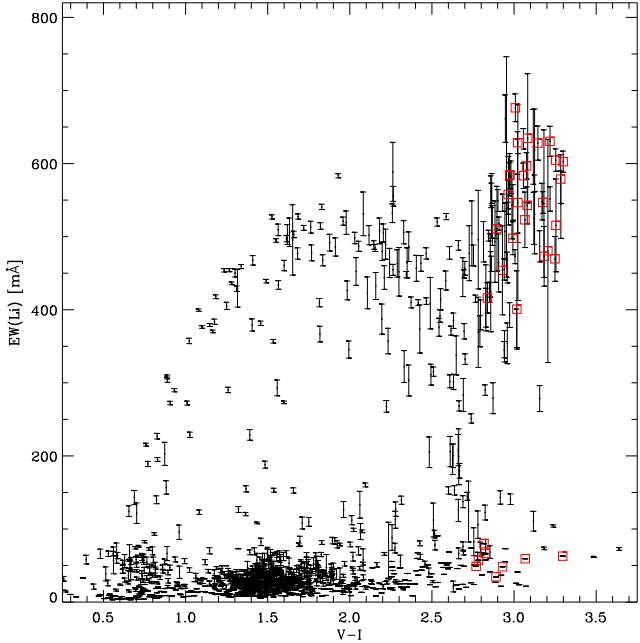}
\caption{The EW(Li) as a function of the color V-I for all targets observed in the $\gamma$ Velorum  region.
Red empty squares are  the fiducial candidate cluster members selected from their RV and the position on the CMD.}
\label{lithiuma2}
 \end{figure}

 Figure\,\ref{lithiuma2} shows  EW(Li) vs.  V-I color where the {\it cluster member fiducial sample},
 selected using only the RVs and the position on the CMD, is highlighted in red. 
 Since this cluster is not affected by  strong reddening, the V-I colors,
 at least for cluster members, can be   considered as a good proxy for the spectral 
 type \citep{jeff09,dami14}.
  We note that, in general, most of the candidate cluster members have EW(Li) larger than 200\,m$\AA$,
 with a trend depending on the spectral type, as expected from the young ages of these objects.
 Nevertheless,  candidate
 cluster members with colors  in the range 2.5$\lesssim$V-I$\lesssim$3,  corresponding
 to stars of spectral type M3 and M4,
 could have a much weaker line and appear to have begun to deplete their Li.
  
We  use the {\it cluster member fiducial sample} 
to empirically define the  cluster locus in this diagram and to distinguish the cluster population
from the field stars. Since the EW(Li) of cluster members shows a pattern that depends on color,
we define four V-I ranges ([1.0-1.5],[1.5-2.0], [2.0-2.5] and [3.0-3.5]) where the EW(Li) distribution
of candidate cluster members is well separated from that of the field stars. This is not the case  
for the bin 2.5$<$V-I$<$3.0, which is treated separately since in this color range,
the EW(Li) of cluster members cannot easily be 
distinguished from those of  field stars. For each of these color ranges,
we assume that the EW(Li) of the candidate cluster members are drawn from an intrinsic
Gaussian distribution that is broadened by   uncertainties on the EW(Li).
For each color range, the {\it cluster member fiducial sample}
includes few contaminants with weak lithium that likely
belong to the field population, so actually we are dealing with two populations.
Therefore we modeled the EW(Li) distribution of the {\it cluster member fiducial sample}
with two Gaussian components, one for the cluster (L$_C$) and one for the field (L$_F$)
to take   account of the small  fraction of contaminants, and fitted the distribution for each
color range 
 using a maximum likelihood technique. In this step, we are only interested 
in the parameters of  the cluster (L$_C$), that
 are given in columns 2 and  3 of 
 Table\,\ref{ewlipar}.
  \begin{table*}
\caption{Parameters  derived with the maximum likelihood fitting for the EW(Li) PDFs.
Column\,1 indicates the color range, cols.\,2 and 3 indicate the  mean and sigma of the cluster
PDF, cols.\,4 and 5 give the mean and sigma of the field PDF while col.\,6 gives the fraction of field
stars with respect to the total sample. Finally, col.\,7 gives the adopted EW(Li) threshold.
  \label{ewlipar}}
\centering
\begin{tabular} {c c c c c c c c}  
\hline\hline
V-I & $<EW(Li)^{Cl}>$ &  $\sigma_{EW(Li)^Cl}$ & $<EW(Li)^{F}>$   & $\sigma_{EW(Li)^Cl}$ & $\frac{N_F}{N_{Tot}}$ & EW(Li)$_{\rm min}$ \\
  & [m$\AA$] &  [m$\AA$] & [m$\AA$] & [m$\AA$] &  & [m$\AA$] \\
\hline
 1.0$<$V-I$<$ 1.5 &422.0& 38.1& 27.9& 18.2&  0.9&100.7\\
 1.5$<$V-I$<$ 2.0 &487.7& 45.1& 32.0& 19.7&  0.9&110.9\\
 2.0$<$V-I$<$ 2.5 &451.4& 58.8& 45.0& 31.7&  0.7&171.7\\
 3.0$<$V-I$<$ 3.5 &555.8& 68.1& 53.3& 27.3&  0.2&162.3\\
\hline
\end{tabular}
\end{table*}

 Next, we considered  all the  targets of the entire dataset for which an EW(Li) value has been released  
 and falling in these color ranges.  
 With the maximum likelihood technique, we fitted again the sum of the two PDFs,
 but in this step, we fixed
 the Gaussian parameters of the cluster L$_C$ to the values
 derived in the first step. The centers and the widths of the  EW(Li) distribution of 
 field stars for each color range, and the fraction of objects that belong to the field population,
 derived in this second step,
 are given in column 4, 5 and 6 of Table\,\ref{ewlipar}.   
 
\begin{figure}
 \centering
 \includegraphics[width=9cm]{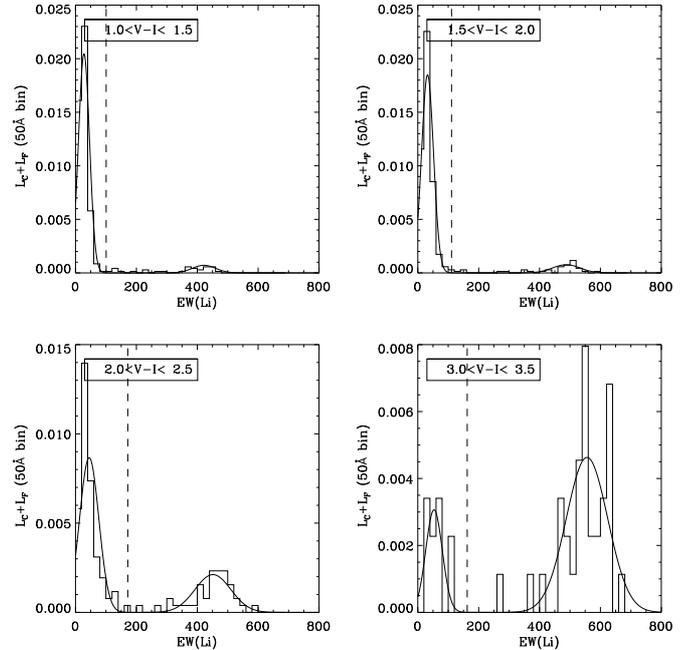}
\caption{Comparison between the  EW(Li) distributions of all observed targets falling in the selected
V-I ranges and the best fit models derived as described in the text. Dashed vertical line in each panel 
indicates the threshold  that has been used to select cluster members.}
\label{lithium_mem}
 \end{figure}

 Figure\,\ref{lithium_mem} shows, for each color range,
 the comparison of the observed EW(Li) distributions from the entire dataset,
with the best fit models derived as described previously.  
We used these models to
derive the best threshold of the EW(Li) to select the maximum number of cluster members whilst minimising the
number of contaminants. For each color range, we define cluster members as those with EW(Li)
 $>4\sigma$ from the mean EW(Li)  of the field PDF L$_F$ (EW(Li)$_{min}$).
By using the field PDF L$_F$, we computed the probability to find contaminants with EW(Li) larger than
 these thresholds (given in column 7 of Table\,\ref{ewlipar}), and then the number of 
contaminants that is $<0.01$. 
Accordingly, with these thresholds, all possible cluster members are expected to be included.

A different approach has been adopted to derive the membership  from the lithium line in the color
range V-I=[2.5-3.0]. Figure\,\ref{lithiuma2} clearly shows that, 
the fraction of  Li-poor fiducial cluster members    
(EW(Li)$\lesssim $100\,m$\AA$) with respect to the number of all observed Li-poor targets   
(21/50=0.42)
in this color range,  is relatively large. 
It is significantly 
higher than the same fractions in the other color ranges, where we find
13/325=0.04, 3/313=0.01 and 4/83=0.05, in the V-I ranges [1.0-1.5],[1.5-2.0] and [2.0-2.5], respectively. 

This suggests that a large fraction of the candidate cluster members 
with very weak lithium and 2.5$<$V-I$<$3.0  are actually cluster members.
Only a small fraction of candidate cluster members, according to their RV,
belong to the field star population. 

To estimate the number of expected cluster members among the 21 Li-poor 
candidates selected for their RV, we need to
estimate the number of expected contaminants.
We hypothesize that outside the range 2.5$<$V-I$<$3,  all the Li-poor stars are unassociated with the cluster.
We further assume that these objects have a similar RV distribution to any contaminating 
field star with $2.5<$V-I$<3$. 
We find that the number of Li-poor stars (considered as contaminants) with $1.0<$V-I$<2.5$ 
selected  within the {\it cluster member fiducial sample}  is 20 (13+3+4) and 
the number of  all observed Li-poor targets in the same color range is 721 (325+313+83).
Then the number
of Li-poor targets not included in the {\it cluster member fiducial sample} is 721-20=701.
Thus the ratio between the contaminants in the {\it cluster member fiducial sample}  and those
outside the {\it cluster member fiducial sample} is 20/701=0.028.
If we assume  the same ratio  in the 2.5$<$V-I$<$3 range, then the number of expected contaminants
in the {\it cluster member fiducial sample}  is 0.028*(50-21)=0.83$\simeq$1.
Therefore, the number  of expected Li-poor cluster members is 21-1=20. 
 For this reason, we cannot rule out that Li-poor targets in this 
 color range are  cluster members.  
 Since we cannot individually assign their membership based on the Li line,
  we consider them as undefined according to Li, leaving  them the chance to be selected
 as cluster members with other membership criteria.
 
Finally, for V-I$<$1, where most of G-type stars are expected to be found,
the strength of the lithium line  is not 
 a sensitive age indicator anymore since these stars do deplete lithium on the Zero Age Main Sequence \citep{sest03}.
 For this reason, in this color range, 
 we consider as  undefined according to the Li  the 14 objects with
EW(Li)$>100$\,m$\AA$, 
while the remaining 154 are considered non members.
We do not consider  the 
4 stars with 3.5$<$V-I$<$5 
and  EW(Li)$<$200\,m$\AA$ as cluster members, since    in this color range they are expected to have
EW(Li)$>$200\,m$\AA$.

After this selection we have 225 
objects with EW(Li) larger than the threshold chosen in each color range,
that are considered cluster members according to  the Li test, 897
non members and
120 objects 
that are undefined according to Li. The last sample includes 
  the 56 objects 
for which the EW(Li) has not been measured,
the 50 objects with EW(Li)$<$100\,m$\AA$ and 2.5$<$V-I$<$3.0, 
and the 14 stars with V-I$<1$ and EW(Li)$>100$\,m$\AA$. 

\begin{figure}
 \centering
 \includegraphics[width=9cm]{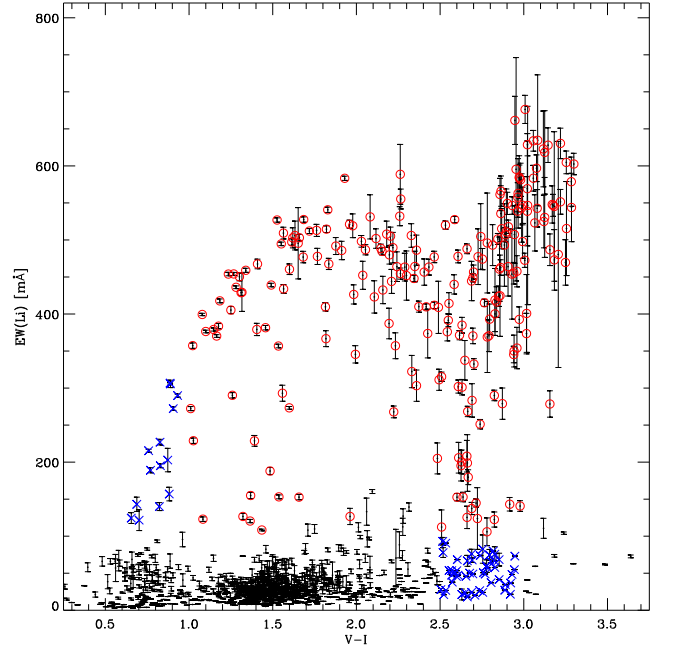}
\caption{The EW(Li) as a function of the color V-I for all targets observed in the $\gamma$ Velorum  region.
Red empty circles are candidate cluster members selected for the Li line criterion, whereas blue crosses are 
the 64  objects which are left undefined according to the Li test.}
\label{lithium_mem2}
 \end{figure}

Figure\,\ref{lithium_mem2} shows the EW(Li) distribution as a function of the V-I colors, where the sample of 
candidate cluster members selected  with the Li line is highlighted.

For  binary stars it is sufficient that one of the two components has an EW(Li)
 larger than the adopted threshold to consider it as a young star. 
However, in the case of candidate binaries, both SB1 and SB2, it is not possible
 to disentangle the  continuum of the two components. In addition, in the case of  
  unresolved SB2 binaries not even  the two lines can be disentangled.
 This implies that the measured EW(Li) can be overestimated or underestimated. 
 Nevertheless,  we considered  the binary stars as single stars,
 with the risk of missing cluster members and/or  including some contaminants,
 This is consistent with our  choice to be inclusive in the selection of candidate members  
 with each criterion taken separately.
\subsection{H$\alpha$ line\label{halphasect}}

Spectra of young stars can show the H$\alpha$ line in emission for several physical reasons, such as
chromospheric activity or accretion of circumstellar material towards  the 
star. This last process can also be associated with outflows from the central star. 
However, while  chromospheric activity affects the core of the line
by filling it and possibly emerging as a narrow  H$\alpha$ emission line, 
accretion and outflow processes affect the line wings causing a significant broadening. 
The H$\alpha$ line broadening arises from the gas motion that implies a
strong enhancement of the gas temperature due to
the shock produced when   the circumstellar material, driven by the magnetic field lines,
 impacts on the stellar surface.
In some case,   also a depression is observed in  the redward wing,  that  is  a signature
of an infalling envelope \citep{bert96}.

A detailed study of the properties of the H$\alpha$ emission profiles for the spectra observed within the Gaia-ESO 
survey  has been presented in \citet{trav15}. Their analysis   highlights several 
morphologic types of the H$\alpha$ emission including the intrinsic emission and the   nebular contribution.


With an age of 5-10\,Myr \citep{jeff09},  the $\gamma$ Velorum cluster could host young stars with accretion, outflows or
chromospheric activity. The H$\alpha$ emission properties from GES spectra
for a sample of selected members of the $\gamma$ Velorum cluster have been extensively studied by
\citet{fras15} who classified accretor stars by using the full width
at 10$\%$ of the H$\alpha$ peak (H$\alpha$\,10\%). In addition, they studied   chromospheric
activity by using the net H$\alpha$
equivalent width derived with a spectral subtraction method \citep{fras94}. This measurement
is based on the  removal of the photospheric flux to obtain the chromospheric emission of the line core.
Their analysis is restricted to the sample of 137 $\gamma$ Velorum members selected as
in \citet{jeff14} with GES spectra having signal to noise ratio (S/N) $>$20.

Based on the previously mentioned properties,
the H$\alpha$ line shape can be used as a membership criterion since it allows us 
to distinguish accretors and young active stars  from non-active older stars.  

In the following sections we describe, starting from the entire GES data set in the $\gamma$ Velorum field,
how we selected spectra with very broadened H$\alpha$ lines,
typical of accretors, and spectra with narrow  H$\alpha$ emission line, characteristic of chromospheric activity.

\subsubsection{Accretor selection \label{accrselec}}

Young stars with accretion are usually selected as  objects with a H$\alpha$\,10\% width $>270$\,km/s
\citep{muze00,whit03,fras15}. By applying this condition to the entire set of GES data in the $\gamma$ Velorum
cluster, we select 26 objects. 
However, since most of the targets observed in the $\gamma$ Velorum  field
are M-type stars and a large fraction of them are also fast rotators, we checked if the broadening observed in the 
H$\alpha$ line occurs also in the other spectral lines, 
rather than in the H$\alpha$ line only, as expected in case of accretion.

 To this aim, we estimated
 the line spectral broadening due to rotation from the FWHM of a rotational (not limb-darkened) line profile, i.e.
\begin{equation}
 \Delta \lambda_{\rm Rot}=2\times\frac{\sqrt3}{2} \frac{vsin~i\lambda_0}{c}
\end{equation}

where $\lambda_0$ is the rest wavelength  and $vsin~i$ is the  projected
rotational velocity.

\begin{figure}
 \centering
 \includegraphics[width=9cm]{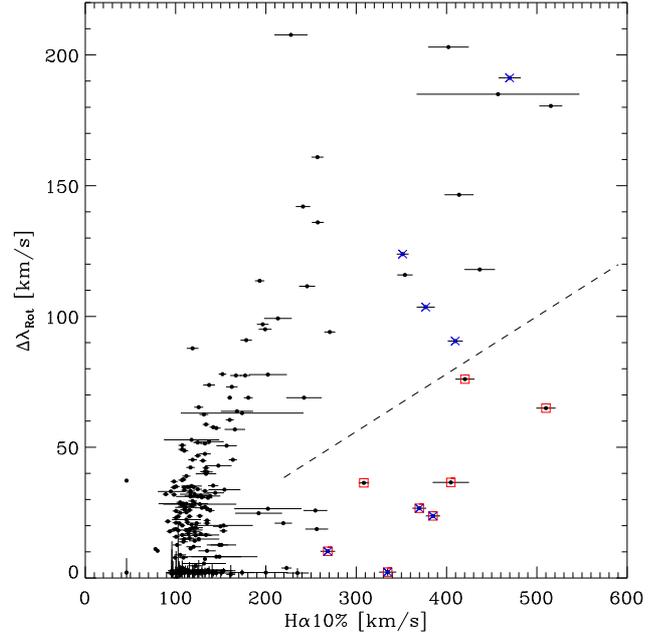} 
 \caption{FWHM of the line spectral broadening due to rotation as a function of the H$\alpha$\,10\%. Empty squares indicate objects classified here as accretors, while crosses indicate the accretors
selected by \citet{fras15}.}
\label{fw10vrot}
 \end{figure}

Figure\,\ref{fw10vrot} shows the $\Delta \lambda_{\rm Rot}$ as a function of the H$\alpha$\,10\%.
It is evident that for a subsample of stars  with large H$\alpha$\,10\% ($>$200\,km/s),  $\Delta \lambda_{\rm Rot}$
is correlated to the H$\alpha$\,10\% and so for these objects the observed broadening of the H$\alpha$
line is likely due to the fast rotation rather than to accretion. 
These objects were not considered  accretors.
  On the contrary, the stars with high H$\alpha$\,10\% but low $\Delta \lambda_{\rm Rot}$
are considered here certain accretors.

In conclusion, we selected as accretors those with H$\alpha$\,10\% larger than 270\,km/s
and $\Delta \lambda_{\rm Rot}$ smaller than the limit (arbitrary chosen) traced by the dashed line 
($\Delta \lambda_{\rm Rot}<0.22\times H\alpha 10\%-10$).
With these conditions, we selected 8 young stars. 

We compared our results with those obtained by \citet{fras15} and we found that 4 of the
8 stars classified here as accretors were also classified by \citet{fras15}. The remaining
4  accretors were not classified   by \citet{fras15} since 
3 of them were not included in their sample and in another case the iDR1 H$\alpha$\,10\%-10 value used 
by \citet{fras15} was 196.5, i.e. smaller than the limit adopted to select accretors.

Finally, there are 4 accretors (CNAME =08083838-4728187, 08094046-4728324, 08104993-4707477 and
08085661-4730350 )
classified by \citet{fras15} that were discarded by us, since their   H$\alpha$\,10\% values
are strongly correlated with the expected rotational broadening and   we suspect that
for these objects the H$\alpha$ line broadening is more related to  fast rotation rather
 than  accretion.

 Since spectra can be variable, especially in case of accretion,
for stars observed more than once, we visually inspected the H$\alpha$ line morphology
using  the single acquired spectra for each target.

We found that both  spectra  of the  star J08075546-4707460  show  a P-Cygni profile, with variable intensity
    in both   emission and absorption components. In addition, the two components are
    correlated in the sense that when the emission intensity decreases, also the absorption  
    decreases.  
    
 In conclusion, we have  8 stars classified as   accretors, including one star with 
   a P Cygni H$\alpha$ profile. These targets are listed in Table\,\ref{accretortable}, where  the objects classified
 by \citet{fras15} are also indicated.
  \begin{table}
\caption{Revised candidate accretor  list in Gamma  Velorum  Column 1 is the CNAME;
column 2 is the FW at 10\% of the H$\alpha$ peak, 
column 3 is the result obtained in this work,
column 4 is the result obtained by Frasca et al. 2015 (FBL15).}
\label{accretortable}
\centering
\begin{tabular}{c c c c }
\hline\hline
Star  & FW10\% & accr. flag & result    \\
      &  km/s &  this work   &    FBL15              \\
\hline
    08065672-4712133  &    404.8$\pm$     20.2   &     Yes  &     No  \\
    08075546-4707460  &    308.4$\pm$      5.9   &Yes-PCyg  &     No  \\
    08082236-4710596  &    510.1$\pm$     10.7   &     Yes  &     No  \\
    08083838-4728187  &    377.0$\pm$     10.1   &      No  &    Yes  \\
    08085661-4730350  &    420.4$\pm$     10.6   &     Yes  &     No  \\
    08094046-4728324  &    469.9$\pm$     12.3   &      No  &    Yes  \\
    08100280-4736372  &    369.9$\pm$      7.7   &     Yes  &    Yes  \\
    08103074-4726219  &    268.5$\pm$      8.2   &     Yes  &    Yes  \\
    08104649-4742216  &    334.8$\pm$      9.5   &     Yes  &    Yes  \\
    08104993-4707477  &    351.4$\pm$      6.6   &      No  &    Yes  \\
    08105600-4740069  &    385.0$\pm$      7.9   &     Yes  &    Yes  \\
    08110328-4716357  &    409.7$\pm$      8.3   &      No  &    Yes  \\
\hline
\end{tabular}
\end{table}

\subsubsection{Active star selection}
Even without accretion activity, young stars with outer convection zones would usually be expected to show 
narrow H$\alpha$ as a result of magnetically-induced chromospheric activity that is ultimately due 
to their relatively fast rotation. Angular momentum loss and spin-down then lead to the fading of
 chromospheric activity with age, but on a mass-dependent timescale - whilst solar-type stars will 
cease to display H$\alpha$ emission on timescale of $\sim 100$ Myr, there can be H$\alpha$ emission
 in lower mass M-dwarfs even at ages of 1 Gyr and beyond \citep{boch07}. Thus narrow H$\alpha$ emission
 lines can be used as a mass-dependent indicator of a youthful status and thus as a condition 
to assign cluster membership in combination with other criteria.

As in \citet{fras15},  to define active stars we considered 
the net H$\alpha$  equivalent width (EWHaChr) values from the GES recommended parameters, available
for 205 of the entire sample of observed stars. In addition,
we used the $\alpha_c$ index
derived by \citet{dami14} that measures the  H$\alpha$ core (2 $\AA$ from the line center) both in cases of emission and absorption.
It has been measured for  1153 stars 
of our sample.
 
\begin{figure}
 \centering
 \includegraphics[width=9cm]{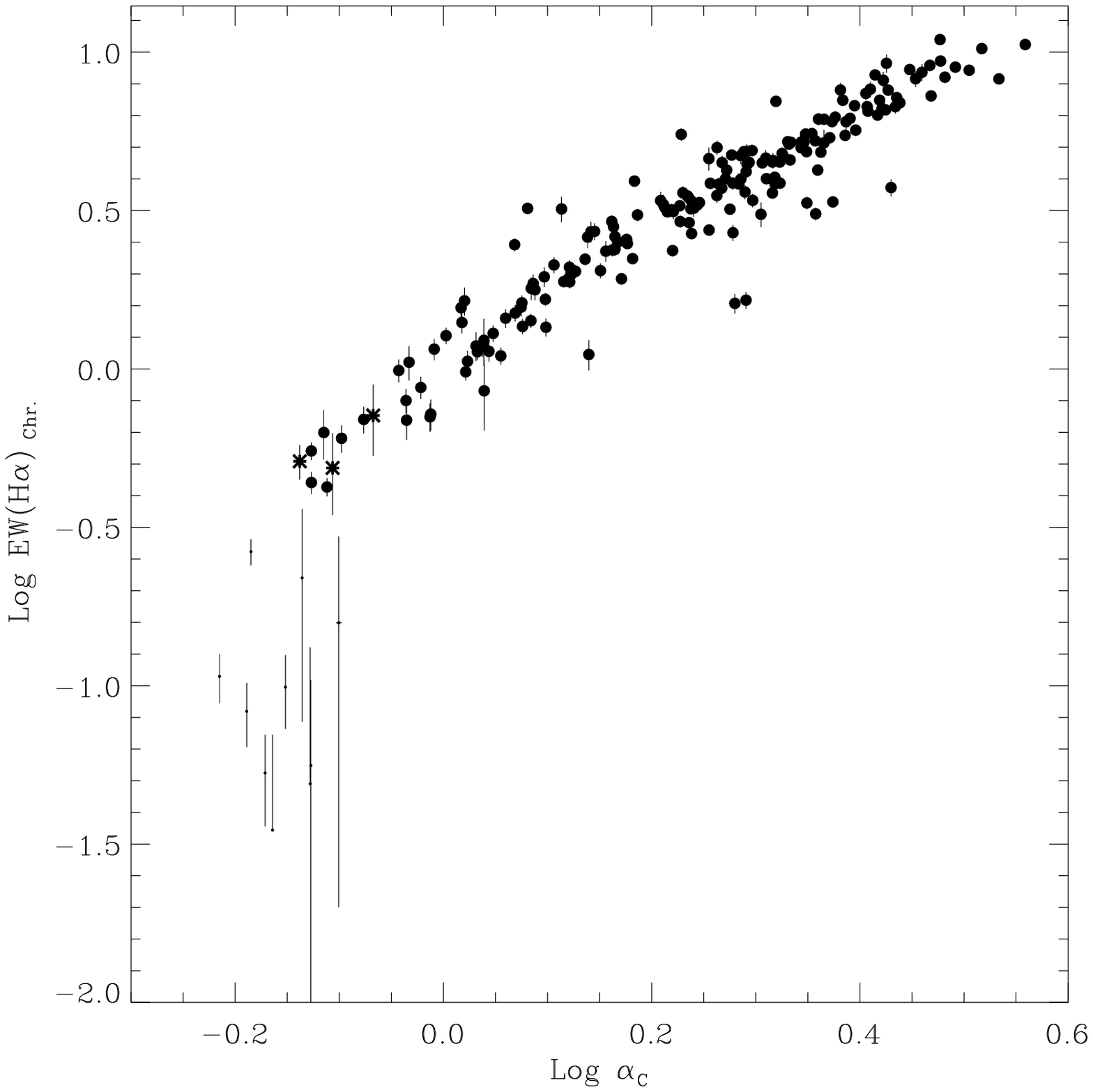}
 \includegraphics[width=9cm]{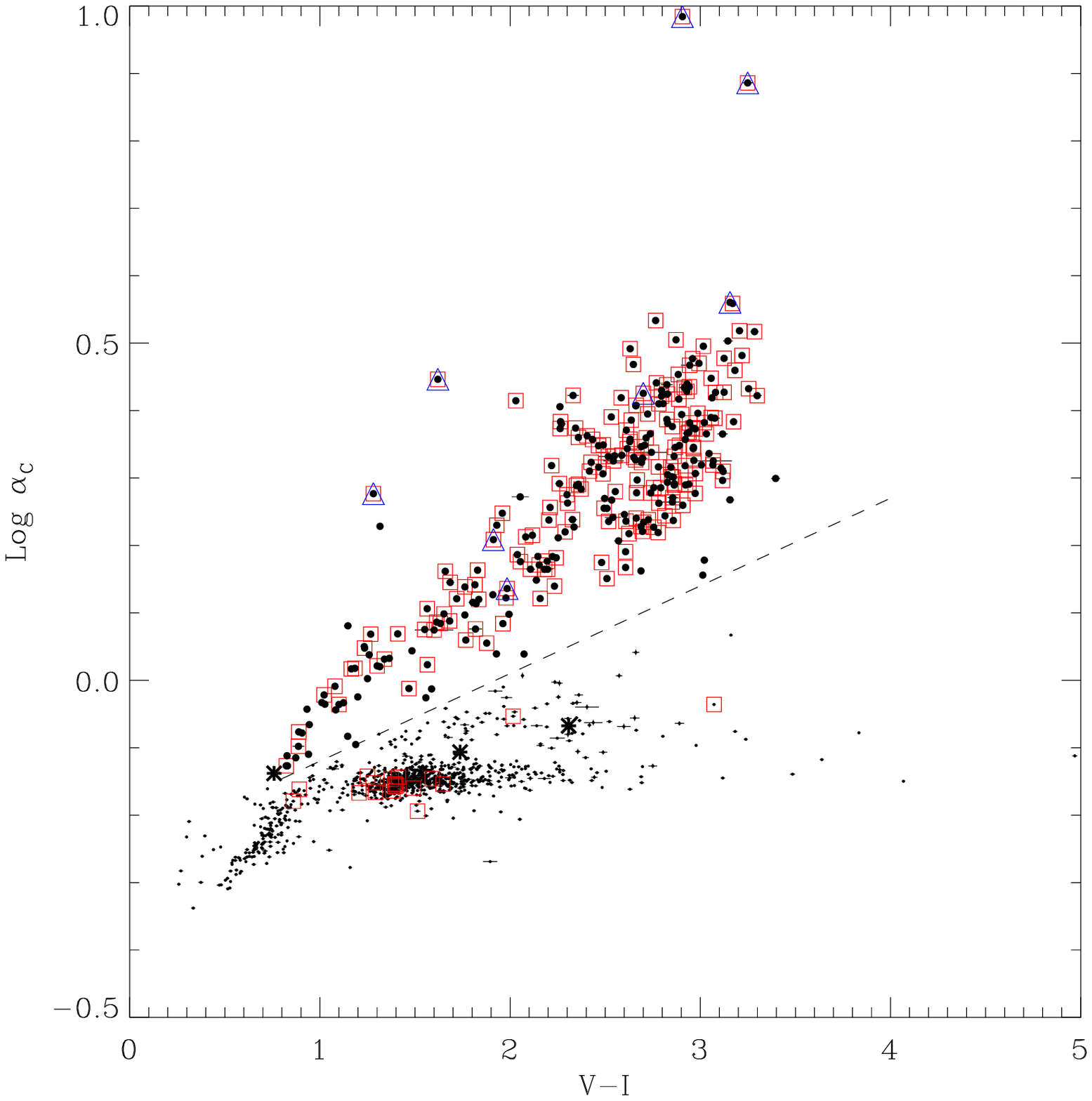}
\caption{Upper panel shows the EW(H$\alpha)_{Chr}$ as a function of  the $\alpha_c$ index while
lower panel shows the  Log $\alpha_c$  index   as a function of the V-I color (dots).
Empty squares are the objects from the {\it cluster members fiducial sample} and
  triangles indicate the objects selected as accretors. 
Filled circles are the active candidate members selected on the basis of the $\alpha_c$ index, 
while asterix symbols are those selected on the basis of the EW(H$\alpha)_{Chr}$. 
The dashed line
indicates the lower limit used for the selection with the $\alpha_c$ index.}
\label{halphamem}
 \end{figure}

Figure\,\ref{halphamem} shows 
the chromospheric EW(H$\alpha$) as a function of  the $\alpha_c$ index (upper panel) and 
the  $\alpha_c$ index as a function of the V-I color (lower panel).
It is evident that, for  Log (EW(H$\alpha_{Chr}))>$-0.5, the chromospheric EW(H$\alpha$)
is well correlated to the  $\alpha_c$ index (upper panel). 
In addition, most of the
cluster members show a characteristic trend for high $\alpha_c$ values as a function of V-I (lower panel), 
that describes the chromospheric emission dependence on spectral type \citep{dami14}. Objects with 
H$\alpha$ absorption line have low $\alpha_c$ values according to the $\alpha_c$ index definition. 

Since the $\alpha_c$ values are given for almost the entire sample of GES observed targets,
we  used this index to select stars with chromospheric activity. In particular,
by following the trend of the $\alpha_c$ index
of the RV candidate cluster members, we define as active stars  the 242 objects with V-I$>$0.8 and
Log\,$\alpha_c > 0.13(V-I)-0.25$ (dashed line) selected from spectra with S/N$>15$. 

The  selected stars correspond to objects with  Log EW(H$\alpha_{Chr})>$-0.5 that 
can  also be considered as a threshold to select confirmed active stars. We discard objects with 
Log EW(H$\alpha_{Chr})<$-0.5 since they 
show very small chromospheric activity and  the EW(H$\alpha_{Chr})$ is affected by large errors.

We added to the sample of selected active members  the 4 objects with 
Log (EW(H$\alpha_{Chr}))>$-0.5 that were not selected in the previous step since their $\alpha_c$ index
is slightly smaller than the threshold we adopted.
In total we selected 246 candidate cluster members on the basis of their chromospheric activity,
10 of which were already  selected as accretors.


\subsection{Candidate members from gravity}
\begin{figure}
 \centering
 \includegraphics[width=9cm]{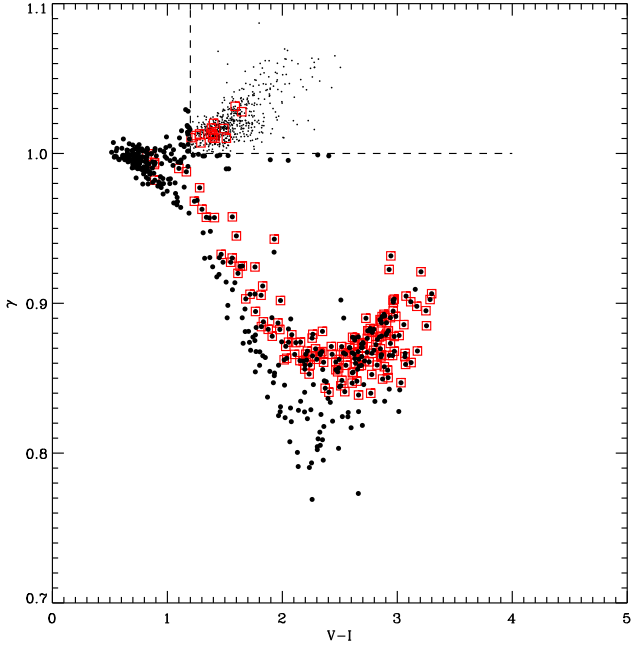}
\caption{Gravity index $\gamma$ as a function of the V-I color (dots). Empty squares are the candidate cluster members selected from their RV and the
position on the CMD  and filled circles
are objects selected as candidate members from gravity. The dashed line
indicates the limit used rejecting giants.}
\label{gravmem}
 \end{figure}

The $\gamma$ index, defined using strongly gravity-sensitive lines  \citep{dami14},
 is an efficient  gravity indicator and 
 allows a clear separation between the low gravity giants and the higher gravity
MS and PMS stars, starting from early G-type stars. Even if with a lower confidence level,
this index allows also to distinguish MS from PMS stars. 
Fig.\,\ref{gravmem} shows the $\gamma$ index as a function of the V-I color for the 1043 objects
for which the index has been released with the GESiDR2iDR3.
Objects with $\gamma \gtrsim 1$ are  giant stars, while those in the bottom region of the plot
are MS and PMS stars.
By using the  {\it cluster member fiducial sample} 
we see that most of them, expected to be PMS stars,
have $\gamma$ index values in the upper
envelope of the region of high gravity objects ($\gamma \lesssim 1$), while MS stars lie in the lower
part of the same envelope. 

We note that this sample does not include the fast rotator stars ($vsini>30$\,km/s)
for which the $\gamma$ index value can be altered by the large line widths \citep{dami14}.

Based on the $\gamma$ index,  
we consider high-probability cluster non members the candidate giants, i.e.  
all the 592 objects with $\gamma> 1.0$ and V-I$>$1.2,
 as indicated by the dashed lines in the Figure. 
These objects correspond to
stars with log\,g$\lesssim 3.2$ and T$_{\rm eff} \lesssim 5600$\,K. 
By using the \citet{sies00} models,
we find that PMS stars with T$<5200$\,K, older than 1\,Myr  have log\,g always greater than $\sim$3.2,
and therefore we are confident that the objects we are discarding are not PMS stars.
We consider all the remaining 648 objects as potential candidate cluster members.

We are aware that by adopting the {\it arbitrary} limit $\gamma=1.0$, we are including a small fraction of
candidate giants with $\gamma\lesssim1.0$ in our sample of candidate cluster members.
This choise is in agreement with  our strategy of being inclusive of all possible candidate cluster members.

This last sample includes the 451 stars that are MS or PMS stars and the 199 objects for which the
gravity index is undefined and for which  membership can be assigned by using the other methods.
We note that with a low confidence level,  MS could be distinguished by PMS stars 
but we adopt the inclusive approach to include in our sample of candidate cluster members even
objects that are MS stars. 

\subsection{X-ray detection \label{xraysection}}
X-ray emission is a further useful criterion  to select cluster members in a young cluster.
 Stellar objects younger than 10$^8$ yrs, such as those expected to belong to the $\gamma$ Velorum cluster,
 are characterized by X-ray fluxes significantly
larger than those observed in older stars of the same spectral type. In particular, in the 0.5-8.0 keV
range, the X-ray luminosity function spans the range between $28< log L_X [erg/s]< 32$, while old solar like 
stars show values 
$26< log L_X [erg/s]< 27$ \citep{fava03,feig07}. This property allows us to distinguish in a very efficient way, 
members in young clusters from field stars expected to be typically older and fainter 
in the X-ray band. The X-ray data can be used here as a membership criterion independent from the spectroscopic
methods discussed before. 

We used here the X-ray catalog compiled in \citet{jeff09} obtained by using two EPIC-XMM-Newton observations 
performed in 2001. Of the 276 individual sources detected considering the two observations,
260 (255 plus additional five sources with optical counterparts with flagged photometry)
have been found in \citet{jeff09} to have an optical counterpart within 6 arcsec, with a
very low fraction of expected spurious matches in the PMS region of the CMD where most of the cluster
members are expected to be found. 

Unfortunately, the XMM-Newton observations cover a field of view of about 30 arcmin in diameter,
where  only 307 of the GES targets fall. Of them, only 106 have an X-ray counterpart in the \citet{jeff09}
catalog. To these 106 sources we added a further 4 targets (CNAME: J08092860-4720178,
           J08093332-4718502,
           J08093364-4722285,
           J08093920-4721387) 
 not included in the \citet{jeff09} X-ray catalog, despite having a clear X-ray counterpart from 
 visual inspection of the  available public EPIC-XMM observations of this field.
 
 In addition, there are 5 X-ray undetected optical sources (CNAME:  
J08092576-4730559, J08093321-4722596,  J08094171-4726420, J08094519-4719061, J08103074-4726219) 
in the  \citet{jeff09} catalog 
  which have an ambiguous X-ray identification, being  
  close to intense X-ray sources or located in region
 with very high background. As in the  \citet{jeff09} catalog, we leave these objects as X-ray undetected
and then we do not consider them as X-ray candidate members.

\begin{figure}
 \centering
 \includegraphics[width=9cm]{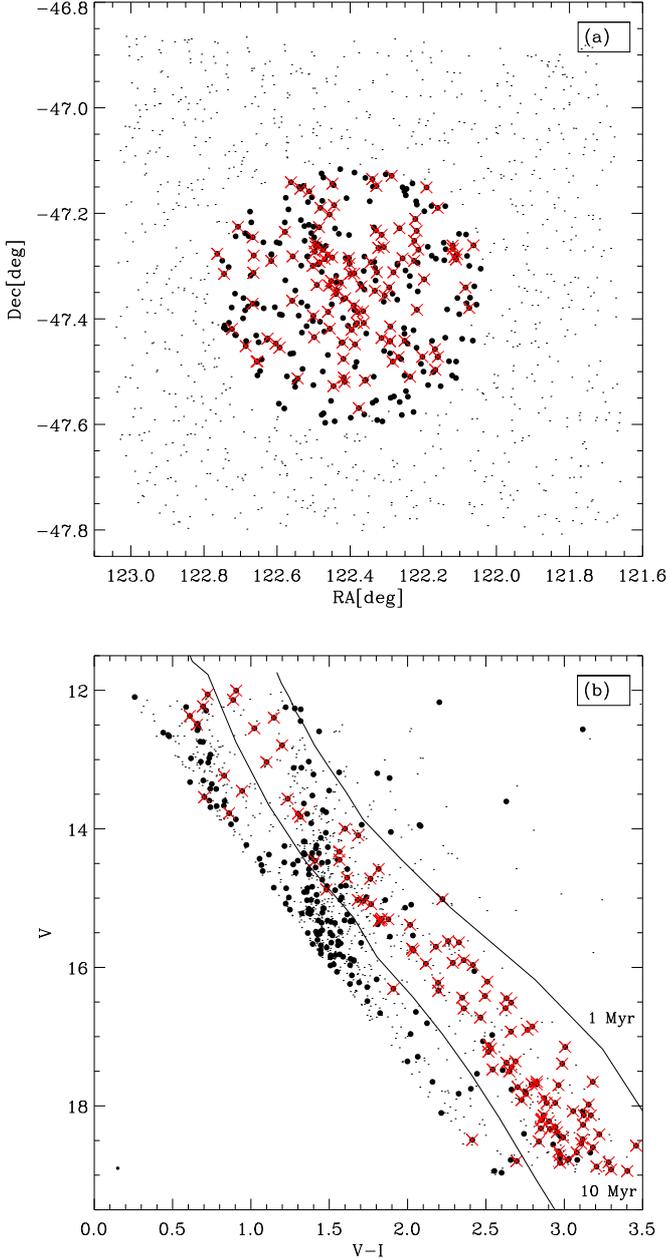}
\caption{Spatial distribution (panel a) and CMD  (panel b) of all GES targets (dots). Filled large circles are all the targets within the EPIC XMM-Newton FOV, while X symbols are the GES targets with an  X-ray counterpart. Solid lines are the 1 and 10\,Myr isochrones by \citet{bara15}.}
\label{xmem}
 \end{figure}

Figure\,\ref{xmem} shows the spatial distribution and the CMD of the 307 targets observed with GES
falling  in the EPIC XMM-Newton field of view (FOV) and the 110 X-ray detections. 
The CMD shows that most of the X-ray detected GES targets  follow the cluster region 
between the 1 and 10\,Myr isochrones, while  the X-ray undetected targets are outside the cluster region.

\section{Final list of members \label{finalmemsection}}
The   membership methods we considered in this work are based on the
spectroscopy  obtained with the GES data, i.e. the RVs, the Li  and H$\alpha$ lines, and the gravity index, and on 
photometry from the literature, i.e. the position of candidates in the CMD
and the X-ray  detections.
In this work we do not consider proper motions since available data are limited to bright stars and
do not help our analysis. 
In addition, we note 
that the S/N limits adopted to define the membership criteria are not the same for all the methods.

As discussed previously, the activity index $\alpha_c$  is derived  by measuring the  H$\alpha$ line core,
while  accretors are defined by measuring the line H$\alpha$\,10\%. This implies that in general
 the sample of active stars includes the accretors, at least when the $\alpha_c$ index is defined,
 and thus we  did not consider here the accretion as a further membership criterion.
We are left with at most 6 independent criteria.

We considered  the gravity index and the photometric criterion  
as  necessary  conditions for cluster membership.
A further necessary requirement for cluster membership is the dynamical 
condition based on the RVs, except for stars identified as binaries and fast rotators
($vsini>50$\,km/s). Indeed, the RVs of these objects  can be affected by the presence of double line series 
(SB2) or by the RV of one of the two stellar components (SB1). In the case of late type fast rotators, the RVs 
are strongly affected by the simultaneous presence of molecular bands and broadening of the spectral lines
due to the rotation. Thus, even in these cases, the RVs can be affected by very large errors and cannot be used 
as a necessary condition to select cluster members.

The other criteria,  i.e. the EW(Li), 
the activity index from the H$_\alpha$ line and the X-ray emission are age indicators and
are used here to confirm the membership\footnote{This choice automatically excludes 
any unidentified short period binaries with RVs outside the cluster RV range.}.

In summary, to  define {\it confirmed members} we required 
that all the following conditions must be fulfilled:
(a) they are members based on their gravity and photometry; (b) they are members for RV; this condition
is not applied to binaries and/or fast rotators; (c)
they are young i.e. they are members based on their Li or   H$\alpha$ index or  X-ray emission.
The conditions (a) and (b) are inclusive of all possible candidates but have the 
disadvantage of also including a fraction of contaminants. However with the condition (c) we are
confident of cutting the contamination very significantly.  
The three youth indicators are sensitive in a different way 
to the spectral types and, in some sense, are complementary, and then they  
are used independently  to ensure  the coverage of the entire spectral type range, especially
where the contamination is worst. In fact,
the Li indicator is most sensitive to age in the K- and M-type objects 
(apart from the narrow window in V-I where Li-depleted M-dwarfs are found),
 but is less effective for G-type stars. On the other hand, the rapid spin-down 
of G-type stars means that X-ray activity is a more effective youth indicator 
in G- and K-type stars, but less effective for M-type stars with their longer 
spin-down and activity timescales \citep[e.g. see discussion in][]{jeff14a}.

We also note that the three age indicators
 have a different sensitivity to the stellar ages. In fact,
depending on the stellar mass, the lithium depletion starts within few million years,
and then very high EW(Li) values allow us to distinguish very young stars. 
The X-ray emission and the chromospheric activity are also decreasing as a function of stellar ages 
but with longer time scale and are very efficient to select low mass stars younger than a few 100\,Myr,
while 
the EW(Li) method is more efficient in selecting stars with ages smaller than $\sim10$\,Myr.

We stress that condition (c) ensure us to include also Li-depleted members with the   very unlikey  risk to include
 unidentified field short period binaries at the same cluster distance and with RV consistent
with that of the cluster.

We note that  we have  optical photometric membership information for the entire data set of
1242 stars, while  the other criteria can be applied only to subsamples.
Table\,\ref{infomemtab}  
  shows the number of objects for which each method can be applied and the corresponding
  number of members by that method. In the case of X-ray detections,
the number of stars for which we have a membership indication is the total number of optical sources falling
in the EPIC-XMM FOV.
\begin{table}
\caption{Number of objects for which we have a membership indication and number of candidate
cluster members for each method
(G=Gravity, P=Photometry, RV=radial velocities, Li=Lithium, A=chromospheric activity, X=X-ray).
\label{infomemtab}}
\centering
\begin{tabular} {c c c }  
\hline\hline
Method & \#info & \#candidates \\
\hline
                                                 G  &  1043  &   451 \\
                                                 P  &  1242  &   579 \\
                                                RV  &  1221  &   541 \\
                                                Li  &  1122  &   225 \\
                                                 A  &  1176  &   261 \\
                                X\tablefootmark{a}  &   307  &   110 \\
\hline
\end{tabular}
\tablefoot{
\tablefoottext{a}{only in the EPIC-XMM FOV}
}
\end{table}


 We started the selection by considering only the sample of the 312
candidates for which both the  photometry and gravity suggest 
membership\footnote{For  spectra with S/N$<15$
we considered only the photometric condition, since the gravity index in these cases is poorly
defined.}.
Among these we considered {\it confirmed members} the 227 objects
with RV compatible with the cluster and
at least one of the three age indicators consistent with young stars. To these we added 15 stars
classified as binaries for which the RV has not been considered but 
that are members by at least for one of the three age indicators. In total we have 242 {\it confirmed members}.
This sample includes 28 fast rotators with RV compatible with that of the cluster. In addition,
we defined  {\it possible members} the 4 fast rotators ($vsini >50$\,km/s)
that are members according to Li or H$_\alpha$ or X-rays, but 
for which the RV is out of the cluster RV  range. As already stressed, for these objects the RVs 
can be unreliable due to the simultaneous presence of molecular bands and line rotational broadening.
All the remaining objects are considered {\it non members}.
  
Table  
\ref{criteriatablecm} summarizes, for the sample of {\it confirmed members},
the six criteria used and the number of cases that we find for each combination.
 \begin{table}
\caption{Criteria adopted to select confirmed members. Abbreviations for the methods
are as in Tab.\,\ref{infomemtab}; (1,0,-) stand for member, non member and no 
information, respectively. M indicates the number of methods for which
the membership is positive while N indicates the number of methods for which
the membership information is available. Finally, the number of cases for each combination is given.
  \label{criteriatablecm}}
\centering
\begin{tabular} {c c c c c c c c c}  
\hline\hline
G & P & RV & Li & A & X & M & N & \#stars \\
\hline
-  &1  &0  &-  &1  &0  &   2  &   4  &   1 \\
-  &1  &1  &1  &-  &-  &   3  &   3  &  11 \\
-  &1  &1  &-  &1  &-  &   3  &   3  &   4 \\
-  &1  &1  &-  &-  &1  &   3  &   3  &   1 \\
-  &1  &1  &0  &1  &-  &   3  &   4  &   2 \\
-  &1  &1  &1  &-  &0  &   3  &   4  &   1 \\
1  &1  &1  &-  &1  &-  &   4  &   4  &  17 \\
-  &1  &1  &1  &1  &-  &   4  &   4  &  19 \\
-  &1  &1  &1  &-  &1  &   4  &   4  &   6 \\
-  &1  &1  &-  &1  &1  &   4  &   4  &   6 \\
1  &1  &1  &0  &1  &-  &   4  &   5  &   4 \\
1  &1  &1  &1  &0  &-  &   4  &   5  &   1 \\
1  &1  &1  &0  &0  &1  &   4  &   6  &   1 \\
1  &1  &1  &0  &1  &0  &   4  &   6  &   1 \\
1  &1  &1  &1  &1  &-  &   5  &   5  &  79 \\
1  &1  &1  &-  &1  &1  &   5  &   5  &   6 \\
-  &1  &1  &1  &1  &1  &   5  &   5  &  17 \\
1  &1  &1  &1  &1  &0  &   5  &   6  &   6 \\
1  &1  &1  &1  &1  &1  &   6  &   6  &  57 \\
\hline
\end{tabular}
\end{table}

\begin{figure}
 \centering
 \includegraphics[width=9cm]{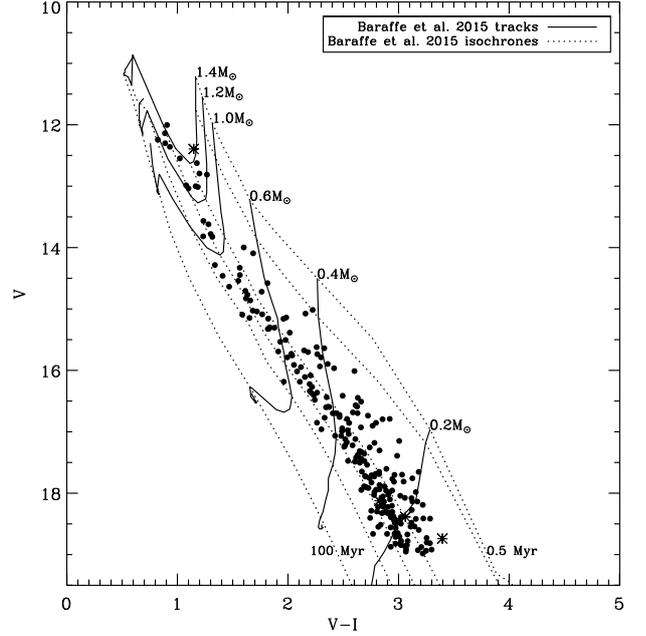}
\caption{Color magnitude diagram of the confirmed (dots) and possible members (crossed dots).
Theoretical tracks and isochrones (0.5, 1, 5, 10, 20 and 100 \,Myr) 
by \citet{bara15} are also shown with solid and dotted lines, respectively.}
\label{massages}
 \end{figure}

The CMD of the confirmed and possible members is shown in Fig.\,\ref{massages} where 
the theoretical tracks and isochrones by \citet{bara15} are also
 drawn assuming the cluster distance modulus 
7.76 mag and E(V-I)=0.055 as in \citet{jeff09}. These models were used  to derive the stellar masses
 that are reported in Table\,\ref{masstablong} together with other fundamental parameters.
The 15 binaries classified as cluster members are treated here as single stars.
\addtocounter{table}{1}
Errors on  masses  were computed by considering the uncertainties in photometry
and the uncertainty in A$_V$  and E(V-I), respectively, for  magnitudes and colors,
starting from the uncertainty in E(B-V) (0.016), estimated in \citet{jeff09}.
Then, we derived  the  masses  corresponding
to the box limits in the CMD defined by these uncertainties.

Since the  \citet{bara15} models are limited to masses smaller than 1.4\,M$_\odot$,
we derived a mass value for 237 of the 246 confirmed and possible cluster members.
This sample includes objects with masses 
between 0.16 and 1.3\,M$_\odot$. 

\section{Discussion}
\subsection{Efficiency of the cluster membership methods}
As stated in the previous section, to define cluster members we required that the stars 
have photometric and dynamic (RV)  properties consistent with that of the cluster.
From this sample we discarded  giants by using the  gravity index and this
allowed us to reduce significantly the fraction of contaminants.

The three age indicators (EW(Li), $\alpha_c$  and X-rays)  
 have been used to confirm the cluster membership.
 In most cases all the three indicators are consistent but we 
have  targets for which only one or  two criteria give us information on their young age.
This can occur for physical reasons, for example if a star already depleted  lithium,
or if a star does not show X-ray emission,  or for observational reasons,
for example if  X-ray sensitivity was not sufficient to detect the object.
For this reason, to confirm the
membership it is sufficient that at least one of the three age indicators is positive.

The results of our membership strategy are given in Table\,\ref{histxmemtab}
for members within the EPIC-XMM FOV, 
for which we have 6 membership criteria and in particular 3 age indicators.
In this table we give the number of confirmed members 
and the number of objects for which we have a membership indication, 
 for each age indicator. The fraction of confirmed members found
 with each method with respect  to the total sample of confirmed members is also given.
 Finally, we counted the number of members we would miss if we did not consider that method.
 The same information is given by splitting the samples in three different color ranges.
 
 The analogous values are given in Table\,\ref{histnoxmemtab} where we consider 
 confirmed members outside of the EPIC-XMM FOV,  
for which we have 5 membership criteria and in particular 2 age indicators.

The lowest efficiency of the EW(Li) method for V-I$\lesssim$1, roughly corresponding to masses 
$>$1\,M$_\odot$, is due to the rapid formation of the radiative core that prevents the Li depletion.
Thus in this spectral range, the EW(Li) is not very effective in selecting young stars.

In general, these results suggest that all the methods are very effective, being positive  for at least 
$\sim$80\% of stars. They are least effective in the regime of M-type stars where some members can be missed.
For this reason it is crucial to use,  in this spectral range, several age indicators.

The efficiency of the EW(Li) is slightly smaller (about 82\%) than the other methods 
 for V-I$>$2.4.  The presence of Li is  an extremely effective age indicator in M-dwarfs since the selected
stars are definitively very young  but, in the narrow colour range 2.5$<$V-I$<$3.0,
where Li can be depleted, this method 
  is ineffective in the sense that some members can be missed.

 The $\alpha_c$  index, signature of H$\alpha$ emission, and the X-ray emission, are not very effective in
 selecting very young stars, since also young field stars of spectral type M can 
show H$\alpha$ and/or X-ray emission.
But,  if on one hand these methods have the disadvantage to include some  contaminants,
on the other hand, with these methods
all potential cluster members can be selected. Members can be missed only for observational limitations
i.e. when    the  S/N of the spectra is $<15$, and then the index cannot be defined, or if they are objects
very close to very strong X-ray emitters (M-type stars   are typically less bright in X-rays) 
or faint objects for which  the X-ray detection probability is low. Spectra with high S/N and/or  
X-ray observations with high spatial resolution   are required to efficiently use these methods.

We find  that within the XMM FOV, the  members not retrieved with the Li line are 16, 
(14 of them are undefined according to Li)
while those not identified  with the H$\alpha$
and X-ray methods are 9,  over a total of 103 members. 
The last column of Tab.\,\ref{histxmemtab} gives the total number of members minus
the number of members recovered by all other methods but independently from the method
indicated in the line. This tells us the number of members that we would miss if we did not use that method.
Thus, within the XMM FOV, the three methods are equivalent and then if we did not 
use one of them we could still select an almost complete sample of members. 

The same is not true
if we consider the results in the region outside the XMM FOV where we note that 13 and 29 members 
would be missed if we did not use the Li or the activity index, respectively. The first group are 
mainly the objects with V-I$>$2.7 that were identified from  their very large EW(Li) that would likely be missed by
the chromospheric activity method since their spectra have S/N smaller than that required, while the latter group 
includes mainly members with $2.5<$V-I$<3.0$ and EW(Li)$<$100\,m$\AA$ (18 objects)

  We note that this is the region where we estimated to find 20 members 
 and  where  we did not discard candidate
members by leaving the objects undefined according to the Li test (see Section\,\ref{lithiumsection}).
Thus, these are  members of the $\gamma$ Velorum cluster according to RV, photometry and gravity, 
that were confirmed by their chromosperic activity.
 Even if we do not have confirmation by the Li line that they are very young members, it is very 
unlikely that they are field stars.

In general the number of members detected by X-rays or from activity is not significantly 
larger than the members found from Li and this suggests to  us that the small differences
among the methods are  related to their detailed  dependence 
on  the spectral range and on the observational strategy. 

The number of members found with the three age indicators can be used to pinpoint 
 any age spread among the  members. In fact,
as discussed in the previous section, the three  methods have also different sensitivity to cluster ages.
An age spread of a few Myr can only be investigated by using Li, at least for the M-dwarfs, while 
 the X-rays and
the chromospheric activity are not really age dependent at these ages. 
Since this cluster is close
to the Vela OB2 association, expected to be relatively young ($<100$\,Myr), we can, in principle, find
more objects selected in X-rays and/or for activity rather than by  Li.  
However, our results suggest that there is no  large age spread among  members since 
the number of members selected by using the Li line is comparable to those selected by using
the X-ray  and the activity methods. Thus we are confident that all selected members originated from 
the same parent molecular cloud.

 \begin{table}
\caption{Breakdown of confirmed members in the XMM FOV. Column\,1 is the method label,
col.\,2 is the number of confirmed members found with that method,
col.\,3 is the number of confirmed members for which we may apply that method,
col.\,4   is the ratio with respect to the total number of confirmed members and
col.\,5 is the number of members we would miss if we didn't consider that method. \label{histxmemtab}}
\centering
\begin{tabular} {c c c c c}  
\hline\hline
Method & \#members & \#info & Fraction & Missed \\
\hline
\multicolumn{3}{c}{entire V-I range    Tot. 103}\\
\hline
   Li  &   87  &   89  &    0.84  &    1 \\
    A  &   94  &   95  &    0.91  &    2 \\
    X  &   94  &  103  &    0.91  &    2 \\
\hline
\multicolumn{3}{c}{ 0.3$<$V-I$<$ 1.1    Tot.   4}\\
\hline
   Li  &    2  &    2  &    0.50  &    0 \\
    A  &    4  &    4  &    1.00  &    0 \\
    X  &    4  &    4  &    1.00  &    0 \\
\hline
\multicolumn{3}{c}{ 1.1$<$V-I$<$ 2.4    Tot.  34}\\
\hline
   Li  &   32  &   33  &    0.94  &    0 \\
    A  &   33  &   34  &    0.97  &    0 \\
    X  &   33  &   34  &    0.97  &    1 \\
\hline
\multicolumn{3}{c}{ 2.4$<$V-I$<$ 4.1    Tot.  65}\\
\hline
   Li  &   53  &   54  &    0.82  &    1 \\
    A  &   57  &   57  &    0.88  &    2 \\
    X  &   57  &   65  &    0.88  &    1 \\
\hline
\end{tabular}
\end{table}

 \begin{table}
\caption{Same as Table\,\ref{histxmemtab} but for the members in the region outside of the XMM FOV
 \label{histnoxmemtab}.}
\centering
\begin{tabular} {c c c c c}  
\hline\hline
Method & \#members & \#info &  Fraction & Missed  \\
\hline
\multicolumn{3}{c}{entire V-I range    Tot. 139}\\
\hline
   Li  &  110  &  116  &    0.79  &   13 \\
    A  &  126  &  127  &    0.91  &   29 \\
\hline
\multicolumn{3}{c}{ 0.3$<$V-I$<$ 1.1   Tot.   4}\\
\hline
   Li  &    1  &    1  &    0.25  &    0 \\
    A  &    4  &    4  &    1.00  &    3 \\
\hline
\multicolumn{3}{c}{ 1.1$<$V-I$<$ 2.4   Tot.  40}\\
\hline
   Li  &   36  &   39  &    0.90  &    1 \\
    A  &   39  &   40  &    0.98  &    4 \\
\hline
\multicolumn{3}{c}{ 2.4$<$V-I$<$ 5.0   Tot.  95}\\
\hline
   Li  &   73  &   76  &    0.77  &   12 \\
    A  &   83  &   83  &    0.87  &   22 \\
\hline
\end{tabular}
\end{table}

\subsection{The IMF}
According to the GES observational strategy, GIRAFFE targets were selected randomly from 
a sample of photometric candidates
while the UVES targets were selected within a specific color range in order to discard F-type candidate members
that are expected to be fast rotators. This implies that while we are able to estimate the completeness of the
sample of confirmed and possible members observed with GIRAFFE, we cannot estimate how complete is the sample
of members selected with  UVES. For this reason, to derive the IMF of the cluster, 
we do not consider  the targets observed with UVES and we
use only the sample of confirmed and possible members observed with GIRAFFE having
masses  between 0.16 and 1.3\,M$_\odot$.

As already mentioned in the Introduction, this cluster includes the two dynamically distinct populations, A and B.
However, according to a KS test, we find that the probability that the two populations have statistically 
indistinguishable mass distributions is 43\%. For this reason, we will not consider these two populations 
separately in the following discussion. 

Starting from the  sample including the $n_{tot}=237$ 
 confirmed and possible members for which we have derived the mass values,
the  observed  IMF has been derived 
 in the linear form 
\begin{equation}
\xi_0(M)=\frac{ dn}{dM}.
\end{equation}

 The mass bins for the IMF were chosen using the condition 
$\Delta log M=0.15$ 
 slightly larger than the typical mass errors. 
We corrected the IMF for incompleteness  by considering for each mass bin the correction factor given by
the ratio between the number of all potential
photometric candidate  members and the number of actually observed targets.
These correction factors $c$ for each mass bin and the values of the corrected IMF
($\xi(M)=c\xi_0(M)$)
are given in Table\,\ref{imftab}.
\begin{table}
\caption{IMF for the Gamma Vel cluster observed with
GIRAFFE. Column 1 gives the mass bin, column 2 gives  the number of stars counted in each mass bin,
column 3 is the correction factor and column 4   gives the IMF values in the linear form.
 \label{imftab}}
\centering
\begin{tabular} {c c c c}  
\hline\hline
Mass & $\Delta N$ & $c$ & $\xi(M)$ \\
\hline
 0.16   -- 0.22    &   54    & 1.16    &    4.05$\pm$    0.55   \\
 0.22   -- 0.31    &   69    & 1.09    &    3.45$\pm$    0.42   \\
 0.31   -- 0.44    &   51    & 1.11    &    1.83$\pm$    0.26   \\
 0.44   -- 0.63    &   28    & 1.06    &    0.68$\pm$    0.13   \\
 0.63   -- 0.89    &   22    & 1.08    &    0.39$\pm$    0.08   \\
 0.89   -- 1.25    &    9    & 1.12    &    0.12$\pm$    0.04   \\
\hline
\end{tabular}
\end{table}

\begin{figure}
 \centering
 \includegraphics[width=9cm]{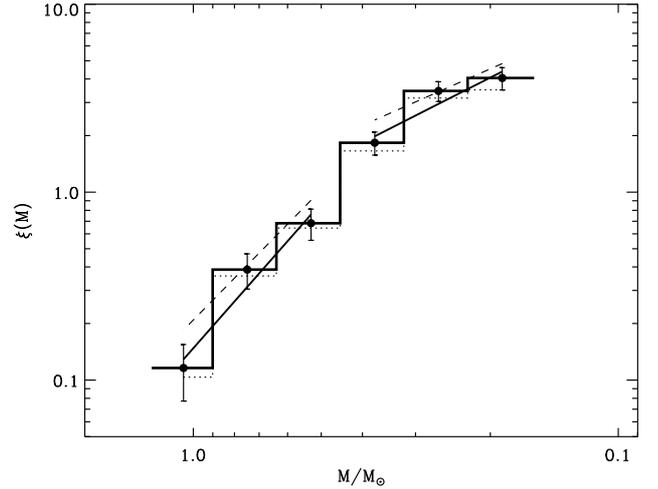}
\caption{The dotted line is the IMF derived from the sample of confirmed and possible members of 
$\gamma$ Velorum  observed with GIRAFFE while the thick solid line is the IMF corrected for incompleteness. 
The overplotted dashed segments represent the \citet{krou01} IMF to which we applied an arbitrary
vertical shift, 
while the solid segments show the IMF obtained with our fit.}
\label{diffimf}
 \end{figure}

The observed and the corrected IMF are shown in 
Fig.\,\ref{diffimf}. 
 We ignore corrections for photometric completeness because they are small; \citet{jeff09} found 
that the level of completeness for stars with good photometry fell only slowly from 93\%
at $V<16$ to 83\% at $19<V<20$.
We note that for all the considered mass bin, the correction factors are 
 $<20$\% and  suggests  that the observed IMF is not very different from the corrected one. 

To derive the  IMF parameters we considered the multiple-part power-law IMF of  stellar populations,
defined by \citet{krou01} in the form 
$\xi(M)\propto M^{-\alpha}$.

We performed a linear fit of the observed IMF and found $\alpha=2.6\pm0.5$ and $\alpha=1.1\pm0.4$ 
in the respective mass ranges. 

The sample of $\gamma$ Velorum cluster members used to derive the IMF includes both the resolved SB1 and SB2 
binaries that we treated as single stars, and an unknown fraction of unresolved binaries. In both cases 
companions are not included in the star-counts. Hence, we compare the observed IMF  with the slopes 
$\alpha=2.3\pm0.5$ for $M>0.5M_\odot$ and $\alpha=1.0\pm0.3$  for $0.15<M/M_\odot <0.5$,
given by \citet{krou13} 
for the primary stars, assuming a binary fraction of 0.5. 
We note that in \citet{krou13}, the slopes of
 the primary IMF are equal to those given for  the canonical IMF of resolved stellar populations,
except in the $0.1<M/M_\odot <0.5$, where the canonical stellar IMF slope is $\alpha=1.3\pm0.3$. 
In Fig.\,\ref{diffimf} we show the results of the linear fit obtained by us compared to the canonical IMF.
This result suggests  that the cluster IMF in the low mass range investigated in this work
is very similar to the canonical one. 


If we consider the mass range used to derive the IMF,
i.e. between 0.16 and 1.3\,$M_\odot$, the total mass of the cluster amounts to $\simeq$92\,$M_\odot$. By considering
the correction for incompleteness, the cluster total mass is $\simeq$100\,$M_\odot$.
Of course this is a lower limit to the total mass, since it is limited to objects with V 
fainter than about 12.5 mag. In addition, we did not consider the binary fraction. Nevertheless,
even by taking into account for the binary fraction and the star component with mass larger than 1.3\,M$_\odot$,
 the observed cluster total mass is hardly compatible with the presence of 
$\gamma^2$ Vel,  whose WC8 component had an initial mass of $\sim 35$\,M$_\odot$.
The expected cluster total mass for a system including a star with  mass  $\sim 35$\,M$_\odot$ is $\sim$1000\,M$_\odot$
\citep{weid10}, significantly larger than the observed one.

Several scenarios have been proposed to explain the formation of $\gamma^2$ Vel and  the surrounding cluster 
\citep{jeff09,jeff14,sacc15}, and
very recently, from N-body modelling,
it has been found that  population A is in virial equilibrium while population B is
strongly supervirial \citep{mape15}. 

Our analysis does not allow to discern between these scenarios  but the finding 
that the entire young population, selected in the region around $\gamma^2$ Vel,
shows a standard IMF suggests us that both  
populations, A and B, formed from the same molecular cloud during the same global star formation process. 




\section{Conclusions}
We have analyzed GIRAFFE
spectra acquired with the GES project and used several   
   membership indicators. This work allowed us to obtain a list
of cluster members whose   membership is confirmed by several criteria
simultaneously. In addition, thanks to the GES target selection strategy, based on an
inclusive sample of candidate members, we were able to obtain a sample of members   
more than 90\% complete.
These achievements are  crucial to study open clusters, usually contaminated by field stars,
for which assessing  membership  is in general very hard.

The GES spectroscopic parameters  used as membership indicators are 
i) radial velocities, 
ii) equivalent withs of the lithium line, 
iii)  $\alpha_c$ index derived by \citet{dami14} from the H$\alpha$
line that gives indications of  chromospheric activity and 
iv) gravity $\gamma$ index defined in \citet{dami14}.  In addition, we used the 
 optical photometry and the X-ray data of the cluster when available.
We obtained a complete list of possible members defined for each method and finally 
a reliable and uncontaminated  as possible list of 246 confirmed members by combining all  information. 

In particular, radial velocities, photometry and gravity index were used as necessary conditions
to select individual stars, while the youth indicators, i.e. lithium, H$\alpha$ and X-ray detections
were used to confirm the membership. For physical reasons or observation limits, 
 youth indicators work best in different spectral regimes. For example,
M-type stars with Li are definitely very young,
even if the lithium depletion 
can occur within few Myr and then stars with this spectral type can show a wide range of lithium abundance.
This implies that by using only the Li criterion, depending on the cluster age,
a fraction of cluster members (about 15\% in our case)
 are missed. On the other hand,
 all the young stars show chromospheric activity or X-ray emission, but, depending on their spectral type,
they are not necessarily very young. Nevertheless, if used in combination with other conditions, 
such as RV, gravity and photometry,   the activity and X-ray criteria are very useful in 
selecting cluster members and allow us to recover also those 
missed according to the Li line in the M-type spectral range.
Since our selection starts from a sample of candidate members for RV, photometry and gravity,
to confirm the membership the three youth indicators were used indifferently.

Finally, by using the new
theoretical models by \citet{bara15}, we derived the masses for 237 of the 246  confirmed and possible members
that are in the range [0.16,1.3]\,M$_\odot$.
 

We derived the cluster IMF by taking into account  the incompleteness due to the unobserved
 members. We compared the derived IMF with the multiple-part power-law IMF form given in \citet{krou01}
and we  found that the IMF slope is  
$\alpha=2.6\pm0.5$   for $0.5<M/M_\odot <1.3$ and $\alpha=1.1\pm0.4$ for $0.16<M/M_\odot <0.5$. 
These values are consistent with a canonical IMF. 

Finally, we found that the total mass of the cluster component with $0.16<M/M_\odot<1.3$ is about 100\,M$_\odot$
that is significantly lower than 
that expected for a cluster in which   a star of $\sim 35$\,M$_\odot$ formed.
The observed IMF suggests us that the two kinematically distinct populations  A and B found by \citet{jeff14},
 formed from the same molecular cloud in the same global star formation process. 
  
\begin{acknowledgements}
This research has made use of data  products from observations made with ESO Telescopes at the La Silla Paranal Observatory under Programme ID 188.B-3002. These data products have been processed by the Cambridge Astronomy Survey Unit (CASU) at the Institute of Astronomy, University of Cambridge, and by the FLAMES/UVES reduction team at INAF/Osservatorio Astrofisico di Arcetri. These data have been obtained from the Gaia-ESO Survey Data Archive, prepared and hosted by the Wide Field Astronomy Unit, Institute for Astronomy, University of Edinburgh, which is funded by the UK Science and Technology Facilities Council.

This work was partly supported by the European Union FP7 programme through ERC grant number 320360 and by the Leverhulme Trust through grant RPG-2012-541. We acknowledge the support from INAF and Ministero dell' Istruzione, dell' Universit\`a' e della Ricerca (MIUR) in the form of the grant "Premiale VLT 2012". The results presented here benefit from discussions held during the Gaia-ESO workshops and conferences supported by the ESF (European Science Foundation) through the GREAT Research Network Programme.
\end{acknowledgements}
\bibliographystyle{aa}
\bibliography{/Users/prisinzano/BIBLIOGRAPHY/bibdesk}

\begin{thebibliography}{31}
\expandafter\ifx\csname natexlab\endcsname\relax\def\natexlab#1{#1}\fi

\bibitem[{{Baraffe} {et~al.}(2015){Baraffe}, {Homeier}, {Allard}, \&
  {Chabrier}}]{bara15}
{Baraffe}, I., {Homeier}, D., {Allard}, F., \& {Chabrier}, G. 2015, ArXiv
  e-prints

\bibitem[{{Bertout} {et~al.}(1996){Bertout}, {Harder}, {Malbet}, {Mennessier},
  \& {Regev}}]{bert96}
{Bertout}, C., {Harder}, S., {Malbet}, F., {Mennessier}, C., \& {Regev}, O.
  1996, \aj, 112, 2159

\bibitem[{{Bochanski} {et~al.}(2007){Bochanski}, {Munn}, {Hawley}, {West},
  {Covey}, \& {Schneider}}]{boch07}
{Bochanski}, J.~J., {Munn}, J.~A., {Hawley}, S.~L., {et~al.} 2007, \aj, 134,
  2418

\bibitem[{{Damiani} {et~al.}(2014){Damiani}, {Prisinzano}, {Micela}, {Randich},
  {Gilmore}, {Drew}, {Jeffries}, {Fr{\'e}mat}, {Alfaro}, {Bensby}, {Bragaglia},
  {Flaccomio}, {Lanzafame}, {Pancino}, {Recio-Blanco}, {Sacco}, {Smiljanic},
  {Jackson}, {de Laverny}, {Morbidelli}, {Worley}, {Hourihane}, {Costado},
  {Jofr{\'e}}, {Lind}, \& {Maiorca}}]{dami14}
{Damiani}, F., {Prisinzano}, L., {Micela}, G., {et~al.} 2014, \aap, 566, A50

\bibitem[{{De Marco} \& {Schmutz}(1999)}]{de-m99}
{De Marco}, O. \& {Schmutz}, W. 1999, \aap, 345, 163

\bibitem[{{de Zeeuw}(1999)}]{de-z99}
{de Zeeuw}, T. 1999, in Astronomical Society of the Pacific Conference Series,
  Vol. 165, The Third Stromlo Symposium: The Galactic Halo, ed. B.~K. {Gibson},
  R.~S. {Axelrod}, \& M.~E. {Putman}, 515

\bibitem[{{Eldridge}(2009)}]{eldr09}
{Eldridge}, J.~J. 2009, \mnras, 400, L20

\bibitem[{{Favata} \& {Micela}(2003)}]{fava03}
{Favata}, F. \& {Micela}, G. 2003, Space Science Reviews, 108, 577

\bibitem[{{Feigelson} {et~al.}(2007){Feigelson}, {Townsley}, {G{\"u}del}, \&
  {Stassun}}]{feig07}
{Feigelson}, E., {Townsley}, L., {G{\"u}del}, M., \& {Stassun}, K. 2007, in
  Protostars and Planets V, ed. B.~{Reipurth}, D.~{Jewitt}, \& K.~{Keil},
  313--328

\bibitem[{{Frasca} {et~al.}(2015){Frasca}, {Biazzo}, {Lanzafame}, {Alcal{\'a}},
  {Brugaletta}, {Klutsch}, {Stelzer}, {Sacco}, {Spina}, {Jeffries}, {Montes},
  {Alfaro}, {Barentsen}, {Bonito}, {Gameiro}, {L{\'o}pez-Santiago}, {Pace},
  {Pasquini}, {Prisinzano}, {Sousa}, {Gilmore}, {Randich}, {Micela},
  {Bragaglia}, {Flaccomio}, {Bayo}, {Costado}, {Franciosini}, {Hill},
  {Hourihane}, {Jofr{\'e}}, {Lardo}, {Maiorca}, {Masseron}, {Morbidelli}, \&
  {Worley}}]{fras15}
{Frasca}, A., {Biazzo}, K., {Lanzafame}, A.~C., {et~al.} 2015, \aap, 575, A4

\bibitem[{{Frasca} \& {Catalano}(1994)}]{fras94}
{Frasca}, A. \& {Catalano}, S. 1994, \aap, 284, 883

\bibitem[{{Gilmore} {et~al.}(2012){Gilmore}, {Randich}, {Asplund}, {Binney},
  {Bonifacio}, {Drew}, {Feltzing}, {Ferguson}, {Jeffries}, {Micela},
  {Negueruela}, {Prusti}, {Rix}, {Vallenari}, {Alfaro}, {Allende-Prieto},
  {Babusiaux}, {Bensby}, {Blomme}, {Bragaglia}, {Flaccomio}, {Fran{\c c}ois},
  {Irwin}, {Koposov}, {Korn}, {Lanzafame}, {Pancino}, {Paunzen},
  {Recio-Blanco}, {Sacco}, {Smiljanic}, {Van Eck}, \& {Walton}}]{gilm12}
{Gilmore}, G., {Randich}, S., {Asplund}, M., {et~al.} 2012, The Messenger, 147,
  25

\bibitem[{{Hern{\'a}ndez} {et~al.}(2008){Hern{\'a}ndez}, {Hartmann}, {Calvet},
  {Jeffries}, {Gutermuth}, {Muzerolle}, \& {Stauffer}}]{hern08}
{Hern{\'a}ndez}, J., {Hartmann}, L., {Calvet}, N., {et~al.} 2008, \apj, 686,
  1195

\bibitem[{{Jackson} {et~al.}(2015){Jackson}, {Jeffries}, {Lewis}, {Koposov},
  {Sacco}, {Randich}, {Gilmore}, {Asplund}, {Binney}, {Bonifacio}, {Drew},
  {Feltzing}, {Ferguson}, {Micela}, {Neguerela}, {Prusti}, {Rix}, {Vallenari},
  {Alfaro}, {Allende Prieto}, {Babusiaux}, {Bensby}, {Blomme}, {Bragaglia},
  {Flaccomio}, {Francois}, {Hambly}, {Irwin}, {Korn}, {Lanzafame}, {Pancino},
  {Recio-Blanco}, {Smiljanic}, {Van Eck}, {Walton}, {Bayo}, {Bergemann},
  {Carraro}, {Costado}, {Damiani}, {Edvardsson}, {Franciosini}, {Frasca},
  {Heiter}, {Hill}, {Hourihane}, {Jofr{\'e}}, {Lardo}, {de Laverny}, {Lind},
  {Magrini}, {Marconi}, {Martayan}, {Masseron}, {Monaco}, {Morbidelli},
  {Prisinzano}, {Sbordone}, {Sousa}, {Worley}, \& {Zaggia}}]{jack15}
{Jackson}, R.~J., {Jeffries}, R.~D., {Lewis}, J., {et~al.} 2015, \aap, 580, A75

\bibitem[{{Jeffries}(2014)}]{jeff14a}
{Jeffries}, R.~D. 2014, in EAS Publications Series, Vol.~65, EAS Publications
  Series, 289--325

\bibitem[{{Jeffries} {et~al.}(2014){Jeffries}, {Jackson}, {Cottaar}, {Koposov},
  {Lanzafame}, {Meyer}, {Prisinzano}, {Randich}, {Sacco}, {Brugaletta},
  {Caramazza}, {Damiani}, {Franciosini}, {Frasca}, {Gilmore}, {Feltzing},
  {Micela}, {Alfaro}, {Bensby}, {Pancino}, {Recio-Blanco}, {de Laverny},
  {Lewis}, {Magrini}, {Morbidelli}, {Costado}, {Jofr{\'e}}, {Klutsch}, {Lind},
  \& {Maiorca}}]{jeff14}
{Jeffries}, R.~D., {Jackson}, R.~J., {Cottaar}, M., {et~al.} 2014, \aap, 563,
  A94

\bibitem[{{Jeffries} {et~al.}(2009){Jeffries}, {Naylor}, {Walter}, {Pozzo}, \&
  {Devey}}]{jeff09}
{Jeffries}, R.~D., {Naylor}, T., {Walter}, F.~M., {Pozzo}, M.~P., \& {Devey},
  C.~R. 2009, \mnras, 393, 538

\bibitem[{{Kroupa}(2001)}]{krou01}
{Kroupa}, P. 2001, \mnras, 322, 231

\bibitem[{{Kroupa} {et~al.}(2013){Kroupa}, {Weidner}, {Pflamm-Altenburg},
  {Thies}, {Dabringhausen}, {Marks}, \& {Maschberger}}]{krou13}
{Kroupa}, P., {Weidner}, C., {Pflamm-Altenburg}, J., {et~al.} 2013, {The
  Stellar and Sub-Stellar Initial Mass Function of Simple and Composite
  Populations}, 115

\bibitem[{{Lanzafame} {et~al.}(2015){Lanzafame}, {Frasca}, {Damiani},
  {Franciosini}, {Cottaar}, {Sousa}, {Tabernero}, {Klutsch}, {Spina}, {Biazzo},
  {Prisinzano}, {Sacco}, {Randich}, {Brugaletta}, {Delgado Mena}, {Adibekyan},
  {Montes}, {Bonito}, {Gameiro}, {Alcal{\'a}}, {Gonz{\'a}lez Hern{\'a}ndez},
  {Jeffries}, {Messina}, {Meyer}, {Gilmore}, {Asplund}, {Binney}, {Bonifacio},
  {Drew}, {Feltzing}, {Ferguson}, {Micela}, {Negueruela}, {Prusti}, {Rix},
  {Vallenari}, {Alfaro}, {Allende Prieto}, {Babusiaux}, {Bensby}, {Blomme},
  {Bragaglia}, {Flaccomio}, {Francois}, {Hambly}, {Irwin}, {Koposov}, {Korn},
  {Smiljanic}, {Van Eck}, {Walton}, {Bayo}, {Bergemann}, {Carraro}, {Costado},
  {Edvardsson}, {Heiter}, {Hill}, {Hourihane}, {Jackson}, {Jofr{\'e}}, {Lardo},
  {Lewis}, {Lind}, {Magrini}, {Marconi}, {Martayan}, {Masseron}, {Monaco},
  {Morbidelli}, {Sbordone}, {Worley}, \& {Zaggia}}]{lanz15}
{Lanzafame}, A.~C., {Frasca}, A., {Damiani}, F., {et~al.} 2015, \aap, 576, A80

\bibitem[{{Mapelli} {et~al.}(2015){Mapelli}, {Vallenari}, {Jeffries},
  {Gavagnin}, {Cantat-Gaudin}, {Sacco}, {Meyer}, {Alfaro}, {Costado},
  {Damiani}, {Frasca}, {Lanzafame}, {Randich}, {Sordo}, {Zaggia}, {Micela},
  {Flaccomio}, {Pancino}, {Bergemann}, {Hourihane}, {Lardo}, {Magrini},
  {Morbidelli}, {Prisinzano}, \& {Worley}}]{mape15}
{Mapelli}, M., {Vallenari}, A., {Jeffries}, R.~D., {et~al.} 2015, \aap, 578,
  A35

\bibitem[{{Muzerolle} {et~al.}(2000){Muzerolle}, {Brice{\~n}o}, {Calvet},
  {Hartmann}, {Hillenbrand}, \& {Gullbring}}]{muze00}
{Muzerolle}, J., {Brice{\~n}o}, C., {Calvet}, N., {et~al.} 2000, \apjl, 545,
  L141

\bibitem[{{Pasquini} {et~al.}(2002){Pasquini}, {Avila}, {Blecha}, {Cacciari},
  {Cayatte}, {Colless}, {Damiani}, {de Propris}, {Dekker}, {di Marcantonio},
  {Farrell}, {Gillingham}, {Guinouard}, {Hammer}, {Kaufer}, {Hill}, {Marteaud},
  {Modigliani}, {Mulas}, {North}, {Popovic}, {Rossetti}, {Royer}, {Santin},
  {Schmutzer}, {Simond}, {Vola}, {Waller}, \& {Zoccali}}]{pasq02}
{Pasquini}, L., {Avila}, G., {Blecha}, A., {et~al.} 2002, The Messenger, 110, 1

\bibitem[{{Pozzo} {et~al.}(2000){Pozzo}, {Jeffries}, {Naylor}, {Totten},
  {Harmer}, \& {Kenyon}}]{pozz00}
{Pozzo}, M., {Jeffries}, R.~D., {Naylor}, T., {et~al.} 2000, \mnras, 313, L23

\bibitem[{{Sacco} {et~al.}(2015){Sacco}, {Jeffries}, {Randich}, {Franciosini},
  {Jackson}, {Cottaar}, {Spina}, {Palla}, {Mapelli}, {Alfaro}, {Bonito},
  {Damiani}, {Frasca}, {Klutsch}, {Lanzafame}, {Bayo}, {Barrado},
  {Jim{\'e}nez-Esteban}, {Gilmore}, {Micela}, {Vallenari}, {Allende Prieto},
  {Flaccomio}, {Carraro}, {Costado}, {Jofr{\'e}}, {Lardo}, {Magrini},
  {Morbidelli}, {Prisinzano}, \& {Sbordone}}]{sacc15}
{Sacco}, G.~G., {Jeffries}, R.~D., {Randich}, S., {et~al.} 2015, \aap, 574, L7

\bibitem[{{Sestito} {et~al.}(2003){Sestito}, {Randich}, {Mermilliod}, \&
  {Pallavicini}}]{sest03}
{Sestito}, P., {Randich}, S., {Mermilliod}, J.-C., \& {Pallavicini}, R. 2003,
  \aap, 407, 289

\bibitem[{{Siess} {et~al.}(2000){Siess}, {Dufour}, \& {Forestini}}]{sies00}
{Siess}, L., {Dufour}, E., \& {Forestini}, M. 2000, \aap, 358, 593

\bibitem[{{Spina} {et~al.}(2014){Spina}, {Randich}, {Palla}, {Sacco},
  {Magrini}, {Franciosini}, {Morbidelli}, {Prisinzano}, {Alfaro}, {Biazzo},
  {Frasca}, {Gonz{\'a}lez Hern{\'a}ndez}, {Sousa}, {Adibekyan}, {Delgado-Mena},
  {Montes}, {Tabernero}, {Klutsch}, {Gilmore}, {Feltzing}, {Jeffries},
  {Micela}, {Vallenari}, {Bensby}, {Bragaglia}, {Flaccomio}, {Koposov},
  {Lanzafame}, {Pancino}, {Recio-Blanco}, {Smiljanic}, {Costado}, {Damiani},
  {Hill}, {Hourihane}, {Jofr{\'e}}, {de Laverny}, {Masseron}, \&
  {Worley}}]{spin14}
{Spina}, L., {Randich}, S., {Palla}, F., {et~al.} 2014, \aap, 567, A55

\bibitem[{{Traven} {et~al.}(2015){Traven}, {Zwitter}, {Van Eck}, {Klutsch},
  {Bonito}, {Lanzafame}, {Alfaro}, {Bayo}, {Bragaglia}, {Costado}, {Damiani},
  {Flaccomio}, {Frasca}, {Hourihane}, {Jimenez-Esteban}, {Lardo}, {Morbidelli},
  {Pancino}, {Prisinzano}, {Sacco}, \& {Worley}}]{trav15}
{Traven}, G., {Zwitter}, T., {Van Eck}, S., {et~al.} 2015, ArXiv e-prints

\bibitem[{{Weidner} {et~al.}(2010){Weidner}, {Kroupa}, \& {Bonnell}}]{weid10}
{Weidner}, C., {Kroupa}, P., \& {Bonnell}, I.~A.~D. 2010, \mnras, 401, 275

\bibitem[{{White} \& {Basri}(2003)}]{whit03}
{White}, R.~J. \& {Basri}, G. 2003, \apj, 582, 1109

\end{thebibliography}
\longtab{5}{
\begin{longtable}{ccccccccccc}
\caption{\label{masstablong} Fundamental parameters of the confirmed and possible members in Gamma  Velorum
Column\,1 is the object name, col.\,2 and 3 give the literature photometry, col.\,4, 5 and  6 are 
the RV, the EW(Li) and the $\alpha_C$ index, respectively; col.\,7 is the X-ray flag where 1 stands for X-ray source
and 0  for X-ray undetected. Column 8 is the $\gamma$ index while col.\,9  gives the masses;
col.\,10 is the binarity flag (0 stands for single stars, 1 for SB1 and 2 for SB2); finally
col.\,11 is the membership flag where CM stands for confirmed member and PM 
stands for possible member.}\\
\hline\hline
CName & V &  V-I & RV    & EW(Li) & $\alpha_c$ &  X & $\gamma$ & Mass &  Bin. & Mem. \\
      &   &      & [km/s]& [m$\AA$] &            &             &          &  [$M_\odot$]    & & \\
\hline
\endfirsthead
\caption{continued.}\\
\hline\hline
CName & V &  V-I & RV    & EW(Li) & $\alpha_c$ &  X & $\gamma$ & Mass &  Bin. & Mem. \\
      &   &      & [km/s]& [m$\AA$] &            &             &          &  [$M_\odot$]    & & \\
\hline
\endhead
\hline
\endfoot
    08064390-4731532  &        17.65    &         2.67    &        16.35 $\pm$        0.38    &        179.65$\pm$      10.000    &         1.982$\pm$       0.014    & (...)  &         0.858$\pm$       0.012    &         0.316$\pm$       0.009    &     0    &   CM    \\
    08064620-4734401  &        18.55    &         2.82    &        19.70 $\pm$        0.89    &         79.90$\pm$       0.000    &                          (...)    & (...)  &                          (...)    &         0.234$\pm$       0.008    &     0    &   CM    \\
    08064823-4735289  &        18.66    &         2.77    &        12.40 $\pm$        0.37    &         48.90$\pm$       0.000    &         2.761$\pm$       0.028    & (...)  &         0.840$\pm$       0.016    &         0.241$\pm$       0.008    &     0    &   CM    \\
    08065228-4712370  &        16.79    &         2.53    &        33.30 $\pm$        0.39    &         26.80$\pm$       0.000    &         1.851$\pm$       0.009    & (...)  &         0.890$\pm$       0.008    &         0.363$\pm$       0.009    &     0    &   CM    \\
    08065469-4657241  &        12.81    &         1.27    &        19.11 $\pm$        0.28    &        454.10$\pm$       2.100    &         1.171$\pm$       0.001    & (...)  &                          (...)    &         1.185$\pm$       0.058    &     0    &   CM    \\
    08065618-4721015  &        18.80    &         3.26    &        16.47 $\pm$        0.90    &        515.40$\pm$      63.498    &                          (...)    & (...)  &                          (...)    &         0.168$\pm$       0.005    &     0    &   CM    \\
    08065672-4712133  &        18.94    &         3.25    &        15.16 $\pm$        0.61    &        469.60$\pm$      30.688    &         7.691$\pm$       0.078    & (...)  &         0.895$\pm$       0.017    &         0.162$\pm$       0.005    &     0    &   CM    \\
\end{longtable}
}

\end{document}